\documentclass[galaxies,article,accept,moreauthors,pdftex,12pt,a4paper]{mdpi}
\lastpage{x}
\doinum{10.3390/------}
\pubvolume{xx}
\pubyear{2014}
\history{Received: 13 February 2014; in revised form: 31 March 2014 / Accepted: 8 April 2014 / \\Published: 2014}
\usepackage{soul,booktabs}

\graphicspath{{fig/}} 

\Title{Mass Distribution in Rotating Thin-Disk Galaxies
 According to Newtonian Dynamics}

\Author{James Q.~Feng * and C. F. Gallo}

\address{%
Superconix Inc., 2440 Lisbon Avenue, Lake Elmo, MN 55042, USA; E-Mail: chuckgallo@comcast.net} 
\corres{E-Mail:~james.q.feng@gmail.com; Tel.:~+1-651-200-6508.}

\abstract{
An accurate computational method is presented for 
determining the mass distribution in a mature spiral galaxy 
from a given rotation curve by applying 
Newtonian dynamics for an
axisymmetrically rotating thin disk of finite size 
with or without a central spherical bulge.
The governing integral equation for mass distribution
is transformed via a boundary-element method
into a linear algebra matrix equation 
that can be solved numerically
for rotation curves with a wide range of shapes. 
To illustrate the effectiveness of this computational method,
mass distributions in several mature spiral galaxies 
are determined from their measured rotation curves. All the surface mass density profiles predicted by our model exhibit approximately a common exponential law of decay, quantitatively consistent with the observed surface brightness distributions.
When a central spherical bulge is present,
the mass distribution in the galaxy is altered in such a way that
the periphery mass density is reduced, while more mass 
appears toward the galactic center.
By extending the computational domain beyond the galactic edge,
we can determine the rotation velocity outside the cut-off radius,
which appears to continuously decrease and to gradually approach 
the Keplerian rotation velocity
out over twice the cut-off radius. 
An examination of circular orbit stability suggests that
galaxies with flat or rising rotation velocities are 
more stable than those with declining rotation velocities
especially in the region near the galactic edge. Our results demonstrate the fact that Newtonian dynamics can be adequate for describing the observed rotation behavior of mature spiral galaxies.
}
\keyword{Newtonian dynamics; disk galaxy; mass distribution; rotation curve}

\begin{document}

\section{Introduction}

Without direct means for accurate measurement, 
mass distribution in galaxies---gravitationally bound assemblies 
of ($10^5$--$10^{12}$) stars---can only be inferred from 
the observable information
according to the known physical laws.
In astronomy, the observable information is usually carried by 
electromagnetic radiation---the light---emitted from 
the visible objects.
The light can be analyzed to provide information about 
the emitting objects, such as their material constituents,
surface temperature, distance, moving velocity, \emph{etc}.
Observations have shown
that many (mature spiral) galaxies share a common 
structure, with the
visible matter distributed in a flat thin disk,
rotating about their center of mass 
in nearly circular orbits 
({\em cf.} the book of Binney and Tremaine \cite{binney87}).
The speed of the circular motion of objects in galaxies 
can be determined from the Doppler shift of light, 
and its plot against the galactocentric distance
is called the rotation curve or circular speed curve.
The measured rotation curve 
has been considered to provide the most reliable information
for deriving the mass distribution in thin-disk galaxies 
\cite{toomre63, sofue01}.

Another independent means for estimating mass distribution 
is based on the luminosity measurements of the galactic surface-brightness profile
by assuming a given (usually constant) mass-to-light ratio,
the validity of which seems to be rather debatable, especially
when it comes to the quantitative calculation of mass distribution.
The darker edge against a brighter bulge background often seen 
from the edge-on view of disk galaxies suggests a varying mass-to-light ratio,
inconsistent with the constant \linebreak mass-to-light ratio assumption.
The findings of Herrmann and Ciardullo \cite{herrmann05} 
show that the value of the mass-to-light ratio of 
the M33 galactic disk can indeed increase by more than a factor of five 
over a radial distance of six times the disk scale length.
Thus, discrepancies often arise between the observed 
rotation curves and that predicted from the mass distributions
following the surface-brightness profile based on Newtonian dynamics,
leading to the so-called ``galaxy rotation problem'' 
still haunting the astrophysical community to the present day
({\em cf.} the book of Freeman and McNamara \cite{freeman06}).
Such discrepancies have often been considered as evidence of 
the luminous disk embedded in a more extensive halo of dark \mbox{matter 
\cite{rubin80, bosma81}}. 
With the concept of dark matter, a constant mass-to-light ratio is usually
assigned to the luminous disk, and 
various parameters are determined for an assumed dark matter halo
to best fit the observed rotation curve.
However, the values of the mass-to-light ratio for the luminous matter are uncertain,
and the fitted functional form for the dark matter halo is often chosen 
arbitrarily 
\cite{broeils92, navarro96}.
On the other hand, 
Palunas and Williams \cite{palunas00} demonstrated that for many galaxies,
the rotation curves 
observed in the optical disk can be successfully described with the
luminous matter having distinct (constant) values of the mass-to-light ratio
without including the dark matter halo.
Thus, whether a dark matter halo exists or whether it is needed for 
describing rotating galaxies becomes questionable. 
The mathematical models involving the dark matter halo have 
at best been still poorly constrained. 
Actually, if not being relied on for quantitatively
calculating the mass distribution,
the typically exponential decaying profile 
of observed surface brightness
in many galaxies 
\cite{freeman70, binney87} 
suggests a likely general structure of 
decreasing (surface) mass density with the galactocentric distance,
which appears qualitatively consistent with that predicted 
from the measured rotation curves
according to Newtonian dynamics for rotating thin-disk 
galaxies 
\cite{feng11}. 

Given a measured rotation curve, 
to derive the mass distribution in a thin-disk galaxy 
requires physical laws that can make the 
connection between the kinematic behavior 
and the locations of matter.
For galactic dynamics, the best-known, well-established physical laws
are Newton's laws of motion and Newton's law of gravity 
\cite{binney87}.
Thus, we focus on
the mass distribution in rotating thin-disk galaxies 
determined from measured rotation curves
according to Newtonian dynamics.
Although theoretically well-established, 
the actual computational efforts in applying Newtonian dynamics to 
thin-disk galaxies appeared to be much more involved
than that for a gravitational system with spherical symmetry, such as
the solar system.
Serious efforts were made for integrating 
the Poisson equation with mass sources distributed on a disk, 
as summarized by 
Binney and Tremaine \cite{binney87}, 
Bratek, Jalocha and Kutschera \cite{bratek08} and 
Feng and Gallo \cite{feng11}, among others.
The solution directly obtained from such efforts 
is usually the (Newtonian) gravitational potential,
which can yield the gravitational force 
by taking its gradient.
In an axisymmetric disk rotating at steady state, 
the gravitational force
(as the radial gradient of gravitational potential)
is expected to equate to the centrifugal force, due to the rotation at
every point.

Unlike the spherically symmetric mass distribution that generates 
the local gravitational force at a given radial position only depending upon
the amount of mass within that radius,
the gravitational force due to a thin-disk mass distribution 
can be influenced by matter, both inside and outside 
that radius.
Thus, the mass distribution in a thin-disk galaxy 
cannot be determined simply by applying Keplerian dynamics, 
which relates the mass within a radial position to the local rotation speed.
In principle, the rotation speed at a radial position is 
mathematically related to
the mass distribution in the entire galactic disk. 
The fact that the brightness in disk galaxies typically decreases 
exponentially with radial distance indicates 
a practical limitation of rotation curve measurements: the
detectable signal must terminate at a finite radial position: the so-called 
``cut-off radius''.
All measured rotation curves terminate at their cut-off radii, 
although sometimes, the cut-off radii may move further out 
with new signal detection and processing technology development.

Among several possible approaches, using Bessel functions 
has been the method of choice for many authors 
\cite{toomre63, freeman70, nordsieck73, cuddeford93, conway00,
jalocha08, bratek08}, 
probably due to the convenience 
in theoretical derivations.
The mathematical formulations with Bessel functions typically 
contain integrals extending to infinity, which 
has become the major practical difficulty when working with
rotation curves that always terminate at finite cut-off radii.
The part of rotation velocity outside the cut-off radius,
although not observable,
must be constructed based on various assumptions, 
to complete the mathematical formulation
(as discussed by Nordsieck~\cite{nordsieck73}, 
Bosma \cite{bosma78}, 
Jalocha, Bratek and Kutschera \cite{jalocha08}, 
Bratek, Jalocha and Kutschera \cite{bratek08}).

To avoid the need of the fictitious part of the rotation curve outside
the cut-off radius, 
an integral equation for a rotating thin-disk galaxy 
with its edge coinciding with the cut-off radius of the rotation curve
can be formulated according to Newtonian dynamics,
consisting of Green's function in terms of the complete 
elliptic integrals of the first kind and second kind 
\cite{feng11}.
With appropriate mathematical treatments, 
the apparent numerical difficulties associated with singularities
in elliptic integrals can be completely removed
when carefully evaluating 
the mathematical limit.
To enable dealing with arbitrary forms of rotation curves and 
mass density distributions, 
the boundary element method for 
solving integral equations is adopted here
using compactly supported basis functions
instead of that extending to infinity, like Bessel functions.
Hence, the finite physical problem domain for a disk with the edge ending 
at a finite radius 
can be conveniently 
considered by solving a linear algebra matrix problem.
Here, in our treatment, we take the cut-off radius as the 
galactic disk edge outside of which the mass density becomes negligible.
Philosophically, where there is matter, there must be a detectable signal;
therefore, no detectable signal must indicate that there is no matter. 
Thus, by ``cut-off radius'', we mean the theoretical cut-off radius, which 
may not be the same as that currently measured in the rotation curve, due to 
technological limitations, but it will be
approached with the continuous improvement of technology.

Following Feng and Gallo \cite{feng11}, in the present work,
we non-dimensionalize the governing equations, such that
a dimensionless parameter,
which we call the ``galactic rotation number'', appears in
the force balance (or centrifugal-equilibrium) equation,
representing the ratio of centrifugal force and gravitational force.
Together with a constraint equation for mass conservation,
the value of this galactic rotation number can be determined as part of 
the numerical solution.
The value of the galactic rotation number can be used for 
calculating the total galactic mass in the disk from 
the measured galactic (cut-off) radius and characteristic rotation velocity.
While Feng and Gallo \cite{feng11} focused mainly on illustrating 
the computational method with a few idealized rotation curves,
herewith, we apply this method to 
the in-depth evaluation of the realistic rotation curves
available in the open literature 
(e.g., the Sofue website \cite{sofuewebsite},
de Blok \emph{et al}. \cite{deblok08}, \emph{etc}.) 
We also extend our method to including the spherical central core and bulge,
to further applications, such as for determining 
rotation velocity beyond the cut-off radius, and so on and so forth.

\section{Mathematical Formulation and Solution Method}

For the convenience of mathematical treatment,
a rotating galaxy is represented by 
a self-gravitating continuum of axisymmetrically distributed
mass in a circular disk with an edge at finite radius $R_g$ 
(beyond which we expect mass density to diminish precipitously to
the inter-galactic level, having inconsequential gravitational effect
on the galactic disk dynamics).
Without loss of generality, we consider 
the thin disk having a
uniform thickness ($h$) with a variable mass density ($\rho$)
as a function of radial distance ($r$) in
galactocentric cylindrical polar coordinates.
In the situation of the thin disk, 
the vertical distribution of mass (in the $z$-direction) is expected to
contribute an inconsequential dynamical effect, 
especially as the disk thickness becomes
infinitesimal. 
In mathematical terms, the meaningful variable here is
actually the surface mass density
$\sigma(r) \equiv \rho(r) \, h$. 
Here, we choose to use the bulk density, $\rho(r)$, for 
its consistency with the common physical 
perception of a thin disk with nonzero thickness $h$.

For steady rotation, there must be a balance 
between the gravitational force and centrifugal force at every point
in the galactic disk.
If the force density on a test mass at ($r$, $\theta = 0$)
generated by the gravitational attraction due to the 
summation (or integration) of a distributed mass density, $\rho(\hat{r})$,
at a position described by the variables of integration ($\hat{r}$, 
$\hat{\theta}$) is expressed as an integral over the entire disk,
with the distance between ($r$, $\theta = 0$) and 
($\hat{r}$, $\hat{\theta}$) given by $(\hat{r}^2 + r^2 - 2 \hat{r} \, r \,
\cos \hat{\theta})^{1/2}$
and the vector projection given by $(\hat{r} \cos \hat{\theta} - r)$, 
the equation of force balance
in a rotating thin disk can be written as (according to Newton's laws):
\begin{eqnarray} \label{eq:force-balance0}
\int_0^1 \left[\int_0^{2 \pi}
\frac{(\hat{r} \cos \hat{\theta} - r) d\hat{\theta}}
{(\hat{r}^2 + r^2 - 2 \hat{r} r \cos \hat{\theta})^{3/2}}\right]
\rho(\hat{r}) h \hat{r} d\hat{r} 
+ A \frac{V(r)^2}{r}
 = 0 \, 
\end{eqnarray}
where all the variables are made dimensionless
by measuring lengths (e.g., $r$, $\hat{r}$, $h$)
in units of the outermost galactic radius, $R_g$,
the disk mass density ($\rho$) in units of
$M_d / R_g^3$, with $M_d$ denoting the total mass in the galactic disk,
and rotation velocities [$V(r)$] in units of the
a characteristic galactic rotational velocity, $V_0$
(usually defined according to the rotation curve of interest).
The disk thickness, $h$, is assumed to be constant and small
in comparison with $R_g$.
The numerical results for surface mass density, $\rho(r) \, h$,
are expected to be insensitive to the exact value of
$h/R_g$,
as long as it remains small.
There is no difference in terms of the physical meaning between
the notations $(r, \theta)$ and $(\hat{r}, \hat{\theta})$;
but mathematically, the former denotes the independent variables in 
the integral Equation 
(\ref{eq:force-balance0}),
whereas the latter, the variables of integration. 
The gravitational force represented as the summation of
a series of concentric rings is described
by the first (double integral) term, while the centrifugal force
by the second term in
Equation (\ref{eq:force-balance0}).

Non-dimensionalizing the force-balance equation
yields a dimensionless parameter,
which we call the ``galactic rotation number'', $A$, expressed as:
\begin{equation} \label{eq:parameter-A}
A \equiv \frac{V_0^2 \, R_g}{M_d \, G} \, 
\end{equation}
where $G$ ($= 6.67 \times 10^{-11}$ [m$^3$/(kg$\cdot$s$^2$)])
is the gravitational constant.
This galactic rotation number, $A$,
simply indicates the relative importance of
centrifugal force \emph{versus} gravitational force.

Equation (\ref{eq:force-balance0}) 
can either be used to 
determine the surface mass density, $\rho(r)\,h$, 
from a given rotation curve, $V(r)$, or \textit{vice versa}.
However, when both $\rho(r)$ and $A$ are unknown, 
another independent equation is needed to keep 
the mathematical problem well-posed.
According to the conservation of mass, 
the total mass of the galaxy disk, $M_d$, should 
stay as a constant satisfying the constraint:
\begin{equation} \label{eq:mass-conservation}
2 \pi \int_0^1 \rho(\hat{r}) h \hat{r} d\hat{r} = 1
\end{equation}
This constraint offers an addition equation for determining the value of 
galactic rotation number $A$.
With Equations (\ref{eq:force-balance0})--(\ref{eq:mass-conservation}),
the mass density distribution, $\rho(r)$, in the disk,
the galactic rotation number, $A$, and the disk galactic mass, $M_d$,
can all be determined from the measured values of 
$V(r)$, $R_g$, $V_0$ and $h$.
On the other hand,
if $\rho(r)$ and $h$ (or $\rho(r)\,h$), as well as $A$ are given, 
$V(r)$ can, of course, be determined from Equation (\ref{eq:force-balance0}).

The integral with respect to $\hat{\theta}$ in Equation (\ref{eq:force-balance0})
is known to be equivalent to:
\begin{eqnarray} \label{eq:elliptic-integral-form}
\int_0^{2 \pi}
\frac{(\hat{r} \cos \hat{\theta} - r) d\hat{\theta}}
{(\hat{r}^2 + r^2 - 2 \hat{r} r \cos \hat{\theta})^{3/2}} 
= 2 \left[\frac{E(m)}{r (\hat{r} - r)} - \frac{K(m)}{r (\hat{r} + r)}\right]
 \, 
\end{eqnarray}
where $K(m)$ and $E(m)$ denote the complete elliptic integrals
of the first kind and second kind,
with:

\begin{equation} \label{eq:m-def}
m \equiv \frac{4 \hat{r} r}{(\hat{r} + r)^2} \, 
\end{equation}
Thus, Equation (\ref{eq:force-balance0}) can be written in a single-integral form:

\begin{equation} \label{eq:force-balance}
\int_0^1 \left[
\frac{E(m)}{\hat{r} - r} - \frac{K(m)}{\hat{r} + r}
\right]
\rho(\hat{r}) h \hat{r} d\hat{r}
+ \frac12 A V(r)^2
 = 0 \, 
\end{equation}
which is more suitable for 
the boundary element type of numerical implementation.

Following a standard boundary element approach 
\cite{sladek98, sutradhar08},
the governing Equations (\ref{eq:mass-conservation}) and
(\ref{eq:force-balance}) 
can be discretized by dividing the one-dimensional
problem domain $[0, 1]$ into a finite number of line segments
called (linear) elements.
Each element covers a subdomain confined by two end nodes,
e.g., element $i$ corresponds to the subdomain
$[r_i, r_{i+1}]$, where $r_i$ and $r_{r+1}$ are nodal values of
$r$ at nodes $i$ and $i+1$, respectively.
On each element, which is mapped onto a unit line segment $[0, 1]$ in
the $\xi$-domain (\emph{i.e.}, the computational domain),
$\rho$ is expressed in terms of the linear basis functions as:
\begin{equation} \label{eq:rho-xi}
\rho(\xi) = \rho_i (1 - \xi) + \rho_{i+1} \xi \, , \quad 0 \le \xi \le 1 \, 
\end{equation}
where $\rho_i$ and $\rho_{i+1}$ are nodal values of $\rho$ at
nodes $i$ and $i + 1$, respectively.
Similarly, the radial coordinate, $\hat{r}$,
on each element is also expressed
in terms of the linear basis functions by
so-called isoparametric~mapping:
\begin{equation} \label{eq:r-xi}
\hat{r}(\xi) = \hat{r}_i (1 - \xi) + \hat{r}_{i+1} \xi \, , 
\quad 0 \le \xi \le 1 \, 
\end{equation}
If the rotation curve, $V(r)$, is given (as from measurements),
the $N$ nodal values of $\rho_i = \rho(r_i)$ are determined by
solving $N$ independent residual equations over the $N - 1$ element
obtained from
the collocation procedure, {\em i.e.},
\begin{eqnarray} \label{eq:force-balance-residual}
\sum_{n = 1}^{N - 1} \int_0^1 \left[
\frac{E(m_i)}{\hat{r}(\xi) - r_i} - \frac{K(m_i)}{\hat{r}(\xi) + r_i}
\right]
\rho(\xi) h \hat{r}(\xi) \frac{d\hat{r}}{d\xi} d\xi 
+ \frac12 A V(r_i)^2
 = 0 \, , i = 1, 2, ..., N \, 
\end{eqnarray}
with:
\begin{equation} \label{eq:mi-def}
m_i(\xi) \equiv \frac{4 \hat{r}(\xi) r_i}{[\hat{r}(\xi) + r_i]^2} \, 
\end{equation}
where $\rho(\xi) = \rho_n(1 - \xi) + \rho_{n+1} \xi$. 
The value of $A$ can be solved by the addition of
the constraint equation:
\begin{equation} \label{eq:mass-conservation-residual}
2 \pi \sum_{n = 1}^{N - 1} \int_0^1
\rho(\xi) h \hat{r}(\xi) \frac{d\hat{r}}{d\xi} d\xi - 1 = 0 \, 
\end{equation}
Thus, we have $N + 1$ independent equations for determining
$N + 1$ unknowns;
the mathematical problem is well-posed.
With appropriate mathematical treatments of the singularities 
arising from the elliptic integrals 
and boundary conditions at $r = 0$ and $r = 1$, 
the set of linear Equations (\ref{eq:force-balance-residual})
and (\ref{eq:mass-conservation-residual})
for $N + 1$ unknowns (\emph{i.e.}, $N$ nodal values of $\rho_i$ and $A$)
can be put in a matrix form
and then
solved with a standard matrix solver, such as
by Gauss elimination 
\cite{press88}.

As in Feng and Gallo \cite{feng11},
the complete elliptic integrals of the first kind and second kind 
in Equation~(\ref{eq:force-balance-residual}) can
be numerically computed with the formulas
\cite{abramowitz72}:
\begin{equation} \label{eq:K-m1}
K(m) = \sum_{l = 0}^4 a_l m_1^l - \log(m_1) \sum_{l = 0}^4 b_l m_1^l
\end{equation}
and:
\begin{equation} \label{eq:E-m1}
E(m) = 1 + \sum_{l = 1}^4 c_l m_1^l - \log(m_1) \sum_{l = 1}^4 d_l m_1^l \, 
\end{equation}
where:
\begin{equation} \label{eq:m1-def}
m_1 \equiv 1 - m = \left(\frac{\hat{r} - r}{\hat{r} + r}\right)^2 \, 
\end{equation}
Clearly, the terms associated with $K(m_i)$ and $E(m_i)$ in
Equation~(\ref{eq:force-balance-residual}) become singular when $\hat{r} \to r_i$
on the elements with $r_i$ as one of their end points.

The logarithmic singularity can be treated by converting the
singular one-dimensional integrals into non-singular
two-dimensional integrals
by virtue of the identities:
\begin{eqnarray} \label{eq:log-integral-identities}
\left\{
\begin{array}{cc}
\int_0^1 f(\xi) \log \xi d\xi = - \int_0^1 \int_0^1 f(\xi \eta) d\eta d\xi
\\
\int_0^1 f(\xi) \log(1 - \xi) d\xi =
- \int_0^1 \int_0^1 f(1 - \xi \eta) d\eta d\xi \,
\end{array} \right . \, 
\end{eqnarray}
where $f(\xi)$ denotes a well-behaving (non-singular) function of $\xi$
on $0 \le \xi \le 1$.

However, a more serious non-integrable
singularity, $1 / (\hat{r} - r_i)$, exists, due to
the term, \mbox{$E(m_i) / (\hat{r} - r_i)$,} in
Equation~(\ref{eq:force-balance-residual}) as $\hat{r} \to r_i$.
The $1 / (\hat{r} - r_i)$ type of singularity is treated by
taking the Cauchy principle value to 
obtain meaningful evaluation 
\cite{kanwal96},
as commonly done with the boundary element method 
\cite{sladek98, sutradhar08}.
In view of the fact that each $r_i$ is considered to be shared by two
adjacent elements covering the intervals $[r_{i-1}, r_i]$ and
$[r_i, r_{i+1}]$, the Cauchy principle value of
the integral over these two elements is given by:

\begin{equation} \label{eq:CPV-def}
\lim_{\epsilon \to 0} \left[
\int_{r_{i-1}}^{r_i -\epsilon} \frac{\rho(\hat{r}) \hat{r} d\hat{r}}{
\hat{r} - r_i}
+ \int_{r_i + \epsilon}^{r_{i+1}} \frac{\rho(\hat{r}) \hat{r} d\hat{r}}{
\hat{r} - r_i}\right]
\, 
\end{equation}
In terms of elemental $\xi$, Equation~(\ref{eq:CPV-def}) is equivalent to:
\begin{eqnarray} \label{eq:CPV-xi}
-\lim_{\epsilon \to 0} \left\{
\int_0^{1 -\epsilon/(r_i - r_{i-1})} \frac{[\rho_{i-1} (1 - \xi) + \rho_i \xi]
[r_{i-1} (1 - \xi) + r_i \xi] d\xi}{
1 - \xi} \right . \nonumber\\
\left .
-\int_{\epsilon/(r_{i+1}-r_i)}^1 \frac{[\rho_{i} (1 - \xi) + \rho_{i+1} \xi]
[r_i (1 - \xi) + r_{i+1} \xi] d\xi}{
\xi} \right\}
\, 
\end{eqnarray}
Performing integration by parts on Equation~(\ref{eq:CPV-xi}) yields:
\begin{eqnarray}
\rho_i \, r_i \log\left(\frac{r_{i+1}-r_i}{r_i-r_{i-1}}\right) 
-\left(
\int_0^1 \frac{d\{[\rho_{i-1} (1 - \xi) + \rho_i \xi]
[r_{i-1} (1 - \xi) + r_i \xi]\}}{d\xi} 
\log(1 - \xi) d\xi \right . \nonumber\\
\left .
+\int_0^1 \frac{d\{[\rho_{i} (1 - \xi) + \rho_{i+1} \xi]
[r_i (1 - \xi) + r_{i+1} \xi]\}}{d\xi} \log \xi d\xi
\right)
\, \nonumber
\end{eqnarray}
where the two terms associated with $\log \epsilon$ 
cancel out each other; the terms with $\epsilon \log \epsilon$ 
become zero at the limit of
$\epsilon \to 0$, and the first term becomes nonzero when the nodes
are not uniformly distributed (namely, the adjacent elements are 
not of the same segment size). 

At the galaxy center $r_i = 0$,

\begin{equation} \label{eq:at_r=0}
\int_{r_i}^{r_{i + 1}} \frac{\rho(\hat{r}) \hat{r} d\hat{r}}{
\hat{r} - r_i} = \int_0^{r_{i + 1}} \rho(\hat{r}) d\hat{r} \, 
\end{equation}
Thus, the $1/(\hat{r} - r_i)$ type of singularity disappears naturally.
However, numerical difficulty can still arise 
if $\rho$ itself becomes singular
as $r \to 0$, e.g., $\rho \propto 1/r$ 
as for the Mestel disk 
\cite{mestel63}.
The singular mass density at $r = 0$ corresponds to a mathematical cusp,
which usually indicates the need for a finer resolution in the physical space. 
To avoid the cusp in mass density at the galactic center,
we can impose a requirement of 
continuity of the derivative of $\rho$ at the galaxy center $r = 0$.
This can be easily implemented at the first node $i = 1$
to demand $d\rho / dr = 0$ at $r = 0$. 
In discretized form for $r_1 = 0$, we simply have:

\begin{equation} \label{eq:rho-1}
\rho(r_1) = \rho(r_2)
\, 
\end{equation}

When $r_i = 1$ (\emph{i.e.}, $i = N$), we are at the end node 
of the problem domain.
Here, we use a numerically relaxing boundary condition
by considering an additional element beyond
the domain boundary covering the interval
$[r_i, r_{i+1}]$, because it is needed to obtain a meaningful 
Cauchy principle value. 
In doing so, we can also assume $r_{i+1} - r_i$ $= r_i - r_{i-1}$,
such that $\log[(r_{i+1} - r_i)/(r_i-r_{i-1})]$ becomes zero,
to simplify the numerical implementation.
Moreover, it is reasonable to assume $\rho_{i+1} = 0$, because it 
is located outside the disk edge.
With sufficiently fine local discretization,
this extra element
covers a diminishing physical space,
such that its existence becomes
numerically inconsequential.
Thus, at $r_i = 1$, we have:
\begin{eqnarray}
\int_0^1 \frac{d\{[\rho_{i} (1 - \xi) + \rho_{i+1} \xi]
[r_i (1 - \xi) + r_{i+1} \xi]\}}{d\xi} \log \xi d\xi \nonumber \\
= (\rho_{i + 1} - \rho_i) \int_0^1 r(\xi) \log \xi d\xi
+ (r_{i+1} - r_i) \int_0^1 \rho(\xi) \log \xi d\xi 
= \rho_i [r_i - \frac32(r_i - r_{i-1})] \, 
\nonumber
\end{eqnarray}
Now that only the logarithmic singularities
are left, Equation~(\ref{eq:log-integral-identities}) can be used for eliminating
all singularities in computing the integrals 
in Equation~(\ref{eq:force-balance-residual}).

\section{Results}

To obtain numerical solutions,
the value of (constant) disk thickness, $h$, must be provided;
we assume $h = 0.01$ out of many possible choices. 
For computational efficiency, 
we distribute more nodes in the regions 
(e.g., near the galactic center and disk edge)
where $\rho$ varies more rapidly. 
Unless the rotation curve has very steep velocity changes 
that need finer discretization with more elements, 
the typical number of non-uniformly distributed nodes, $N$, 
used in computating most cases
is $1001$ (corresponding to $1000$ linear elements), which we found for most cases to be sufficient for
obtaining a smooth curve of $\rho$ \emph{versus} $r$ and 
discretization-insensitive values of galactic rotation number $A$.

The rotation curves 
available in the open literature 
(e.g., the Sofue website \cite{sofuewebsite})
are typically provided in a tabular form with data points 
at radial positions
often not coinciding with our nodal positions. 
We use the cubic spline interpolation method 
\cite{press88} 
to evaluate our nodal values of $V(r)$ from the rotation curve data,
such that the rotation curve used in our computations 
is guaranteed to smoothly pass through all the original data points.

\subsection{NGC 4736}

The {\em NGC 4736} galaxy has recently been studied by 
Jalocha, Bratek and Kutschera \cite{jalocha08},
for illustrating that the baryonic matter distribution 
can account for the observed rotation curve.
Thus, we believe it deserves our attention to study 
using our computational method. 

There are several different versions of rotation curve data for {\em NGC 4736} 
in the literature.
Here, we consider two of them,
one is from the website of Sofue  
\cite{sofuewebsite}
and the other from THINGS 
(namely, The H I Nearby Galaxy Survey)
 measurements \citep{deblok08}.
Figure 1 shows the two versions of the rotation curves 
with $r$ measured in units of $R_g = 10.35$ (kpc) 
(where $1$ kpc $= 3.086 \times 10^{19}$ m),
and rotation velocity $V(r)$ in units of $V_0 = 150$ (km/s).

As shown by Feng and Gallo \cite{feng11}, the value of total galactic mass 
in the disk can be determined according to Equation~(\ref{eq:parameter-A}) 
with the computed value of $A$ as:

\begin{equation} \label{eq:M_d}
M_d
= \frac{V_0^2 \, R_g}{A \, G} \, 
\end{equation}
Because the computed value of the galactic rotation number, $A$, is 
$1.9656$ for the THINGS rotation curve and $1.5908$ for that of Sofue,
we obtain $M_d = 2.756 \times 10^{10}$ $M_{\odot}$
(where one solar-mass \mbox{$M_{\odot}$ $= 1.98892 \times 10^{30}$~kg)} 
when the THINGS rotation curve is used and 
$M_d = 3.405 \times 10^{10}$ $M_{\odot}$ when the Sofue rotation curve
is used. 
The value of 
$M_d = 3.405 \times 10^{10}$ $M_{\odot}$
agrees well with that computed by 
Jalocha, Bratek and Kutschera \cite{jalocha08} 
(\emph{i.e.}, $3.43 \times 10^{10}$ $M_{\odot}$) using 
the same rotation curve of Sofue. 
\begin{figure}[H]
\centering
{\includegraphics[scale=0.60]{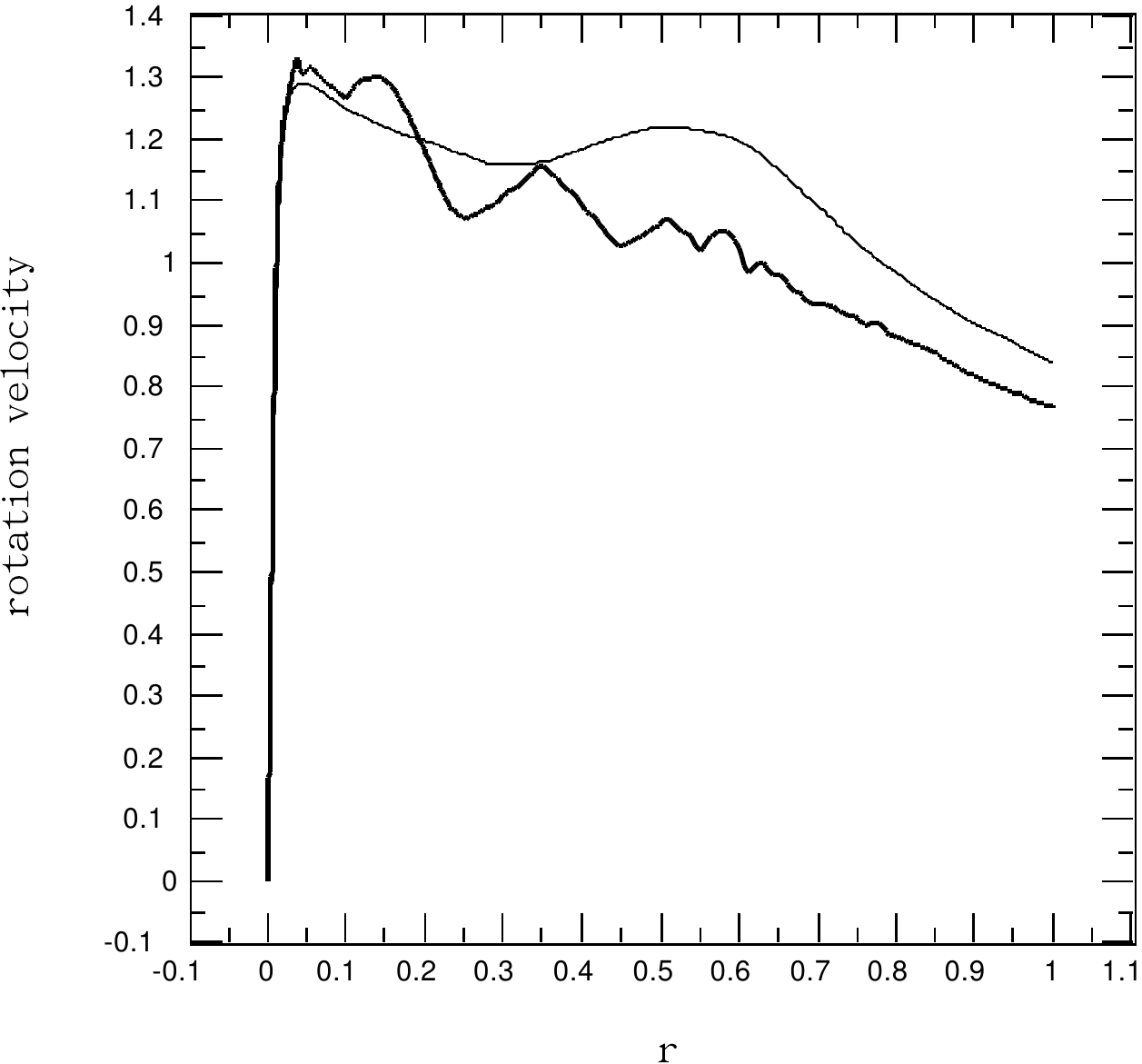}}
{\includegraphics[scale=0.60]{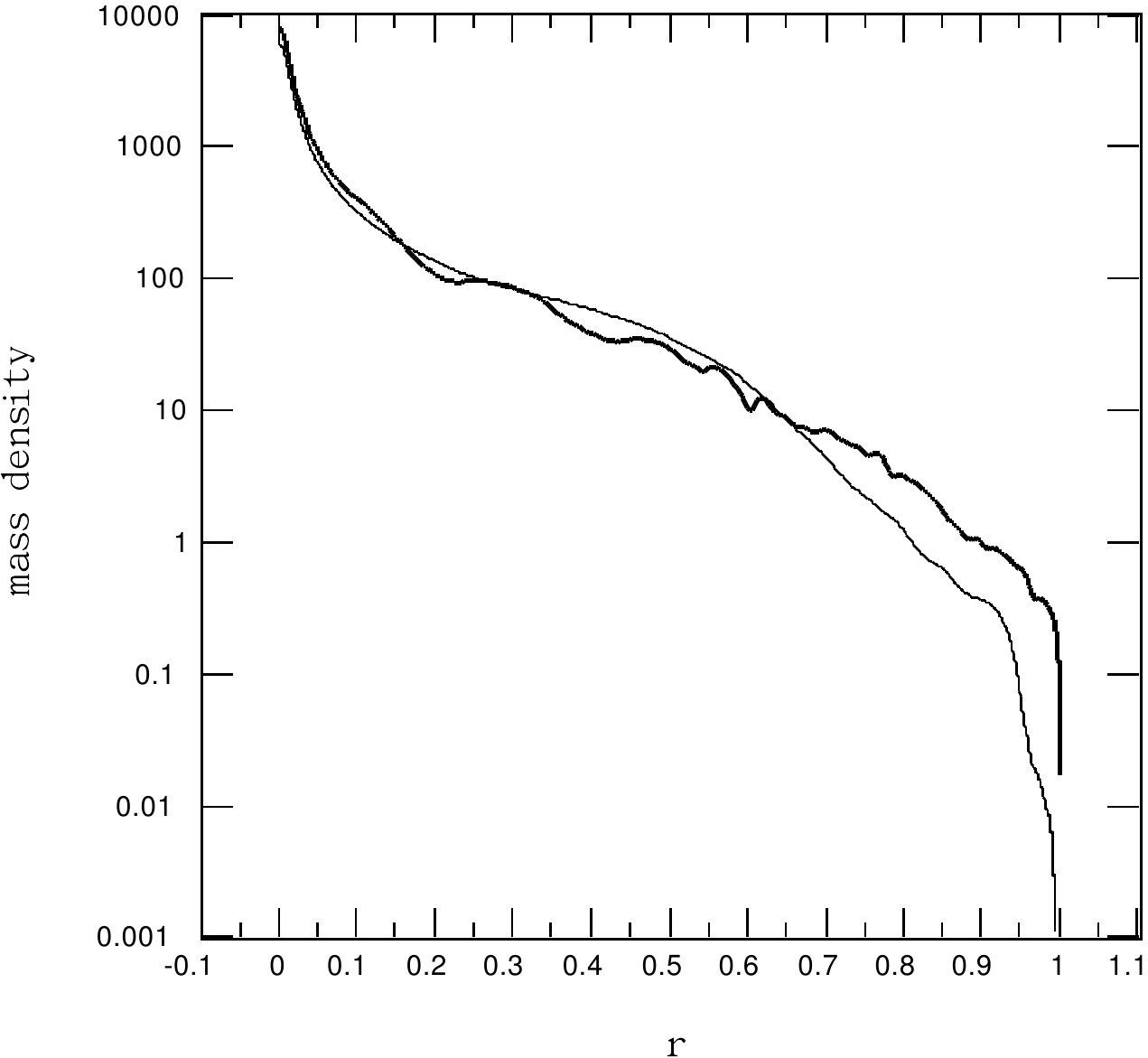}}
\caption{\label{fig:fig1} Profiles of {\em NGC 4736} rotation velocity
$V(r)$ and mass density $\rho(r)$,
with the thick line for that of 
THINGS (The H I Nearby Galaxy Survey)
 and the thin line for 
that from Sofue.
The computed values of the galactic rotation number, $A$,
are $1.9656$ and $1.5908$ for the cases of THINGS and Sofue,
respectively.}
\end{figure}

However, Jalocha, Bratek and Kutschera \cite{jalocha08} 
used an iterative spectral method 
with Bessel functions,
which requires the inclusion of a rotation curve beyond the
``cut-off'' radius extending to infinity.
They also considered the mass density due to 
the neutral atomic hydrogen (H I) 
outside the cut-off radius.
With our method, only the available data for rotation curve 
within the cut-off radius is needed.
Meanwhile, we assume the mass density in 
the galactic disk diminishes at the same cut-off radius
to enable a self-consistent consideration of 
the mathematical problem on a finite disk domain.
The solution of the axisymmetric mass distribution
in the galactic disk
for a given rotation curve can be computed
by one-step Gauss elimination of the linear algebra matrix equation
without further successive \mbox{iterations 
\cite{feng11}.}

If desired, the effect of H I
and the ``not-yet'' measurable rotation curve outside the cut-off radius
can be conveniently examined in an {\em a posteriori} manner.
For example, the surface mass density of H I
considered by Jalocha, Bratek and Kutschera \cite{jalocha08} 
at the cut-off radius
is $\sim 1$ $M_{\odot}/$pc$^2$, 
which translates to our 
non-dimensional $\rho = R_g^2/(M_d \, h) = 0.3887$ (or $0.3146$,
with $R_g$ in units of pc and $M_d$ in units of $M_{\odot}$)
for the THINGS (or Sofue) rotation curve,
decreasing one order of magnitude in about $3$ (kpc) 
beyond the cut-off radius at $10.35$ (kpc).

To examine the effect of H I outside the cut-off radius,
we modify the mass distribution starting from a radius $r_1 < 1$, such that
the mass density for $r \ge r_1$
is described by:
\begin{equation} \label{eq:rho1}
\rho(r) = \rho_1 e^{-[0.2 + (r - r_1) / 0.3]^2}
\, , \, r_1 \le r < \infty \, 
\end{equation}
where $r_1 = 0.965$ and $\rho_1 = \rho(r_1) = 0.388$.
The profile of mass density distribution extending to 
$r > 1$ as described by (\ref{eq:rho1}) 
approximates well to that considered by 
Jalocha, Bratek and Kutschera~\cite{jalocha08}
while simplifying the analysis.
With the given mass distribution extending beyond $r = 1$,
we can correspondingly extend the integration to $r > 1$ 
in Equations~(\ref{eq:mass-conservation}) and (\ref{eq:force-balance}),
to calculate rotation velocity beyond the cut-off radius.
Figure 2 shows the mass density distribution with (thin line)
and without (thick line) the H I outside 
the (non-dimensional) cut-off radius
$r = 1$,
and the corresponding rotation curves.
The integration result of Equation~(\ref{eq:mass-conservation})
shows that including H I beyond 
$r = 1$
increases the total galactic mass only by $\sim 0.5\%$. 
Therefore, it is not surprising to notice that
the original rotation curve 
in Figure 2 is barely altered 
by this mass density modification.
In fact, the rotation curves beyond $r = 1$
calculated with and without the mass density modification
(\ref{eq:rho1}) differs so little
that they are visually indistinguishable when plotted together in Figure 2. 
\begin{figure}[H]
\centering
{\includegraphics[scale=0.60]{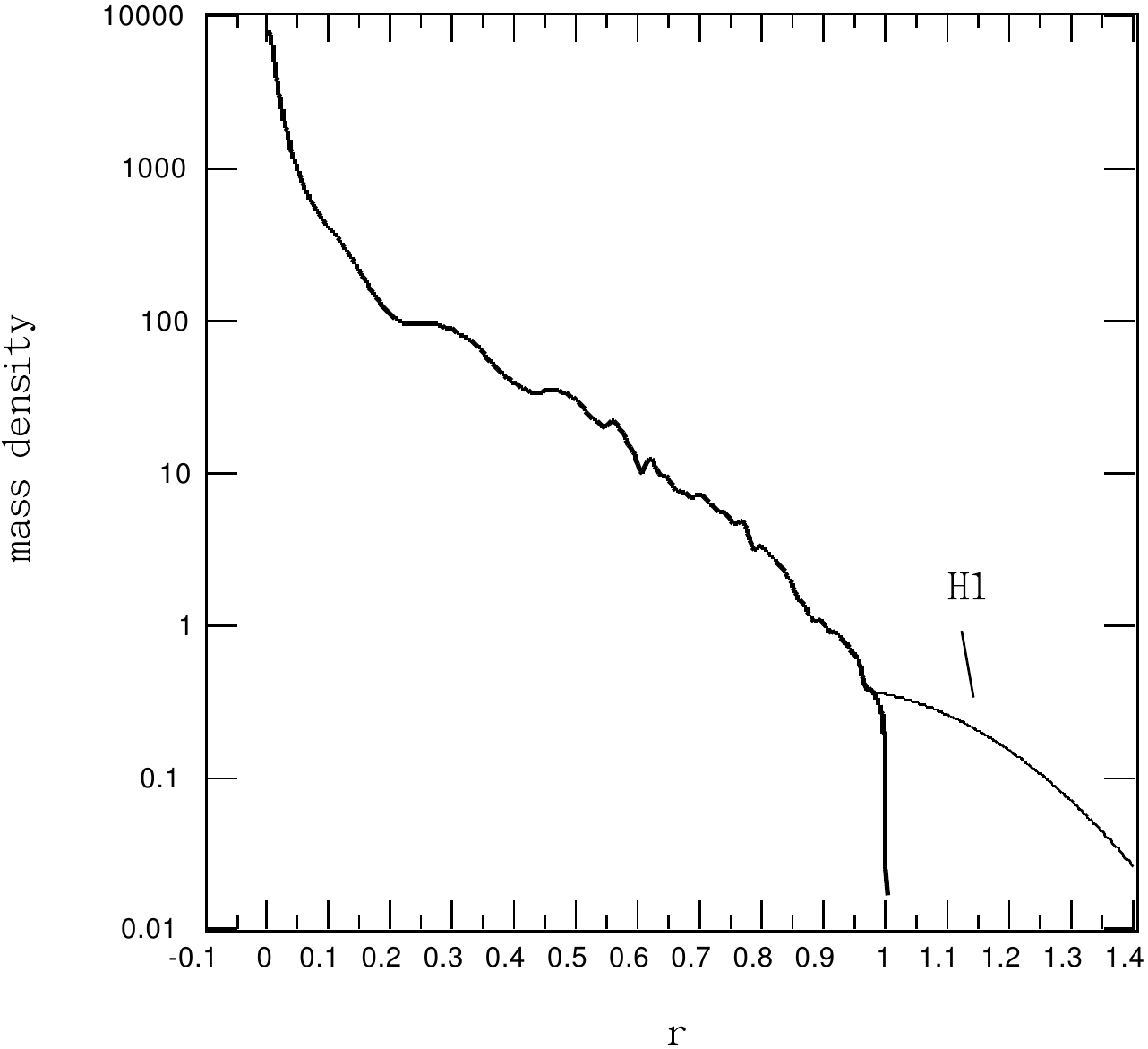}}
{\includegraphics[scale=0.60]{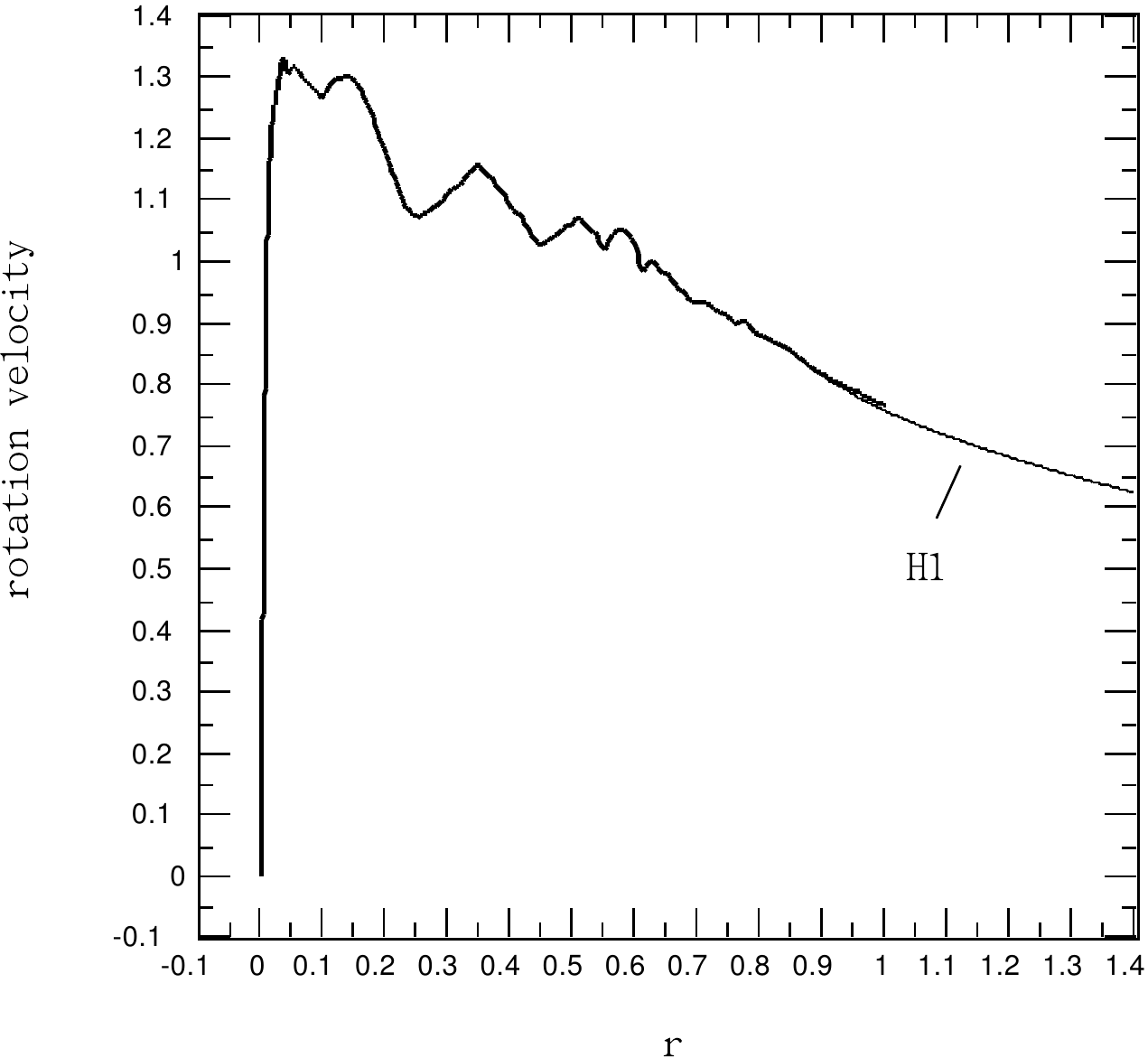}}
\caption{\label{fig:fig2}The distributions of
mass density $\rho(r)$ with (thin line) or without (thick line) 
H I extending beyond the cut-off radius,
and the corresponding rotation curves.
Only the case based on the THINGS rotation curve is shown here.} 
\end{figure}
In view of typical uncertainties in rotation curve measurements
(as illustrated in Figure 1 for two different versions of 
the same galaxy),
we expect that any matters (such as H I) 
outside the ``cut-off'' radius cannot 
have substantial influence on the disk rotation characteristics,
because the amount of mass in comparison to the value of $M_d$ 
is usually insignificant.
Thus, whether including the H I mass outside the 
``cut-off'' radius of {\em NGC 4736} should have inconsequential effect
on the Newtonian dynamics relating the measured rotation curve
to mass distribution in the galactic disk. 
As illustrated here, however, the consideration of H I mass beyond $r = 1$
can be conveniently implemented as an {\em a posteriori} process 
(without iteratively computing solutions) to
evaluate (or, in other words, predict) the rotation velocity beyond
the cut-off radius, 
which could not be obtained from measurements.
Then, the needed part of rotation curve beyond the cut-off radius, 
for using a formulation that requires it 
\linebreak(e.g., \cite{nordsieck73, bosma78, jalocha08, bratek08}) 
to determine the mass distribution from measured rotation curve,
can be provided using our method 
with concrete certainty (namely, without fictitious assumptions).

\subsection{Milky Way, NGC 4945}

The Milky Way, also called the Galaxy, 
is the galaxy that contains the Sun and the Earth,
which is why it is of particular interest to astronomy and astrophysics.
{\em NGC 4945} is a spiral galaxy that appears quite similar to
the Milky Way. 
Therefore, we present results for both of them together here. 
The rotation curve data provided by Sofue 
\cite{sofuewebsite}
suggest nonzero rotation velocity at $r = 0$.
According to our continuum treatment of a rotating disk galaxy
with a Newtonian dynamics description of force balance 
(\ref{eq:force-balance}), 
nonzero rotation velocity at $r = 0$ requires 
a strongly singular mass density to ensure that:

\begin{equation} \label{eq:nonzero_integral}
\int_0^1 \rho(\hat{r}) d\hat{r} \to \infty
\, \, 
\end{equation}
This is because the kernel of integral in 
Equation~(\ref{eq:force-balance}) for any nonzero $\hat{r}$ has a limit value of zero
at \mbox{$r = 0$, \emph{i.e.}, }

\begin{equation} \label{eq:kernel}
\lim_{r \to 0}\left[\frac{E(m)}{\hat{r} - r} - \frac{K(m)}{\hat{r} + r}\right]
=\frac{1}{\hat{r}} \left[E(0) - K(0)\right]
\, \, 
\end{equation}
and $E(0) = K(0) = \pi/2$ \citep{abramowitz72}.
Thus, $V(0)$ must be zero according to 
Equation~(\ref{eq:force-balance}), 
unless Equation~(\ref{eq:nonzero_integral}) is true 
as in Mestel's disk \cite{mestel63} 
where $\rho \to 1/r$ as $r \to 0$.

As discussed by Feng and Gallo \cite{feng11}, 
the computational method used here can reproduce the result of
Mestel's disk \citep{mestel63} for the entire problem domain $(0, 1]$ 
when the rotation velocity in
an infinitesimal neighborhood around $r = 0$ is modified, such that 
$V(0)$ becomes zero,
which corresponds to replacing $\rho(0) = \infty$ with a 
finite (large) value of $\rho(0)$. 
Such a slight modification of rotation curve results in
no practical difference in the computed mass density distribution
and the value of total galactic mass, $M_d$,
while providing great convenience for numerical computation.

Therefore, we take the same approach here to slightly modify
the rotation-curve data files of Sofue, such that the first
point at $r = 0$ has $V(0) = 0$, while leaving all the rest of the 
data points unchanged; the resulting rotation curves are 
shown in Figure 3
with $r$ measured in units of $R_g = 20.55$ and $20.00$ (kpc), 
rotation velocity $V(r)$ in units of $V_0 = 220$ and $180$ (km/s),
respectively for the Milky Way and \mbox{{\em NGC 4945.}}
Also shown in Figure 3 are the computed 
mass density distributions.
With the computed value of $A = 1.6365$, 
we have $1.4138 \times 10^{11}$ $M_{\odot}$
for the Milky Way according to Equation~(\ref{eq:M_d}).
For the Galaxy, one unit of non-dimensional $\rho$ corresponds to 
the surface mass density of $M_d \, h/R_g^2 = 3.35$ $M_{\odot}/$pc$^2$.
In the solar neighborhood around 8 kpc from the Galactic center,
which corresponds to $r = 0.3893$, 
\linebreak we have $\rho \sim 43$ (from Figure 3), and therefore, the surface mass density
around the Sun should be \linebreak$\sim 144$ $M_{\odot}/$pc$^2$.

Due to the large central peaks in rotation curves near $r = 0$,
the computed mass density profiles show a sharp increase of 
$\rho$ toward the galactic center, 
as is consistent with the previous findings of Feng and Gallo~\cite{feng11} 
based on a series of idealized rotation curves.
Because the rotation curves of the Milky Way and {\em NGC 4945} 
are generally flat, 
the mass density profiles show only about a one order of magnitude 
decrease in the large interval $(0.1, 0.9)$,
unlike that for {\em NGC 4736}, with more than a two orders of magnitude 
decrease, corresponding to a rotation curve of velocity generally decreasing 
with galactocentric distance.

\begin{figure}[H]
\centering
{\includegraphics[scale=0.60]{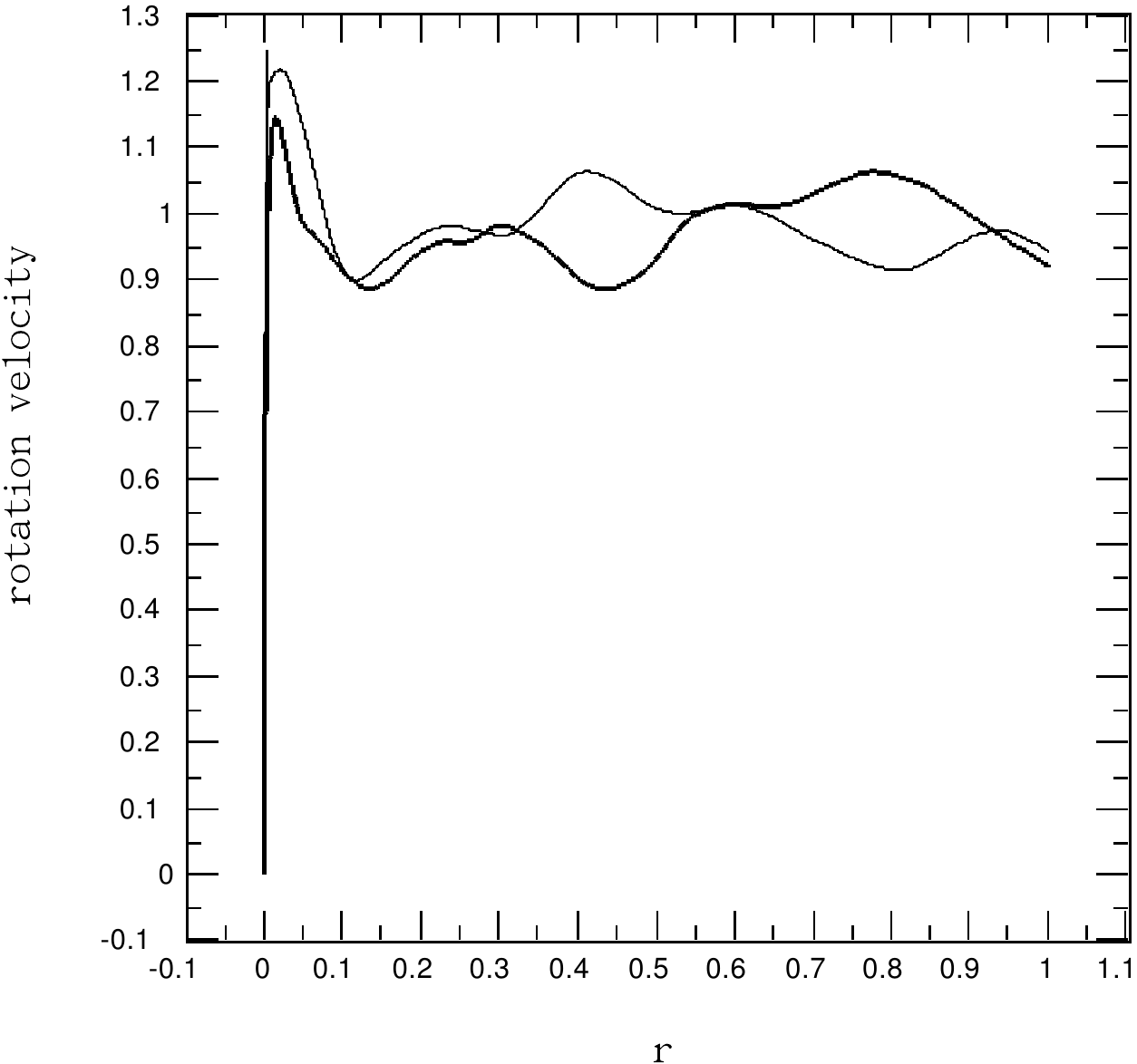}}
{\includegraphics[scale=0.60]{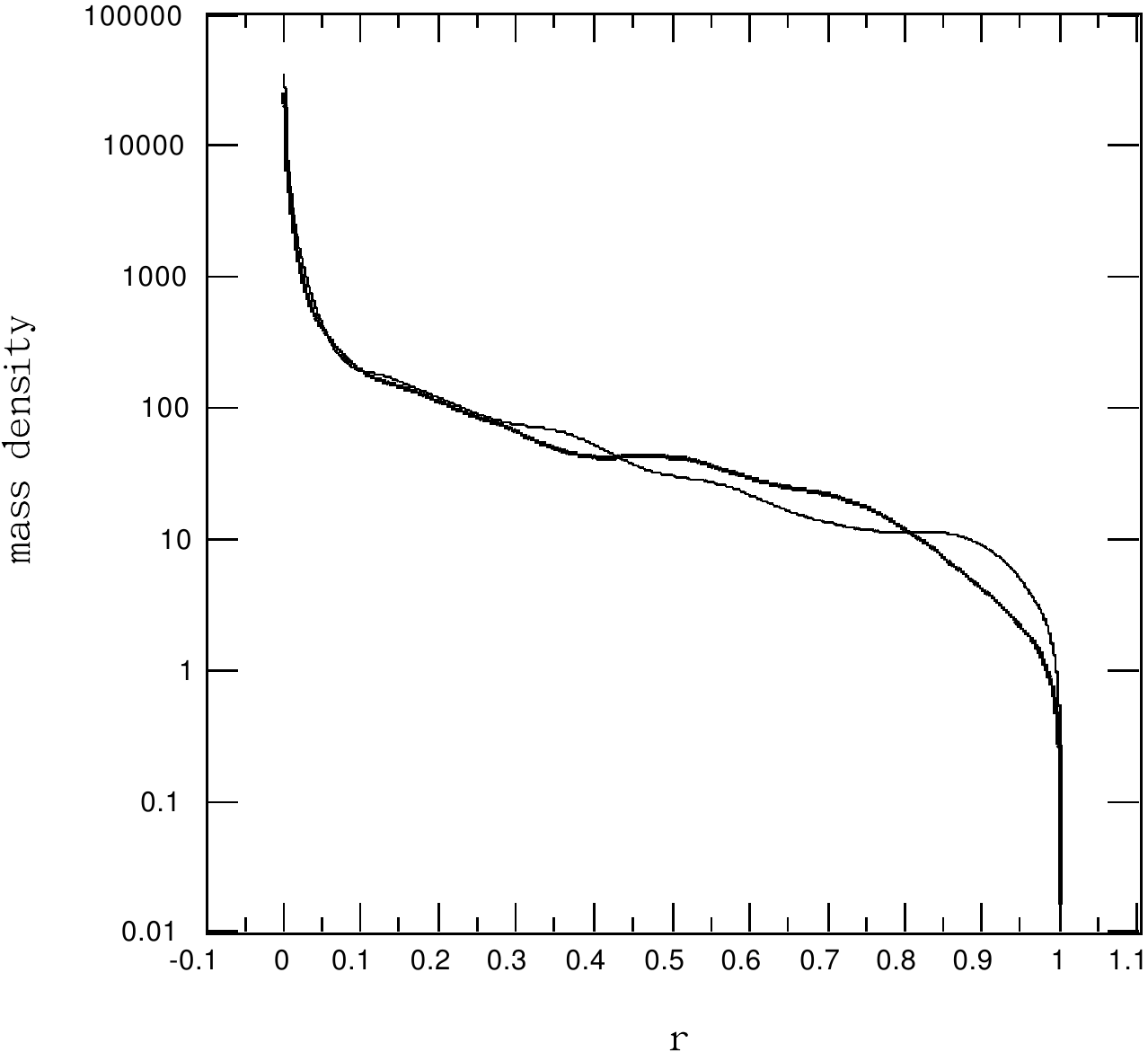}}
\caption{\label{fig:fig3} Profiles of the Milky Way and 
{\em NGC 4945} rotation velocity
$V(r)$ and mass density $\rho(r)$,
with the thick line for that of the Milky Way and the thin line for 
{\em NGC 4945}.
The computed values of the galactic rotation number, $A$,
are $1.6365$ and $1.6873$ for the Milky Way and {\em NGC 4945},
respectively.}
\end{figure}

\subsection{NGC 224, NGC 5055}

The {\em NGC 224} and {\em NGC 5055} galaxies were classified as 
those with rotation curves having ``no central peak'' 
\cite{sofue99}, 
in contrast to that of the Milky Way.
Their rotation curves 
(from the Sofue website \cite{sofuewebsite})
and our computed mass density profiles are 
presented in Figure~4,
with $r$ measured in units of $R_g = 31.25$ and $39.35$ (kpc), 
rotation velocity $V(r)$ in units of $V_0 = 250$ and $190$ (km/s),
respectively for {\em NGC 224} and {\em NGC 5055}.
Corresponding to the rotation curves without the central peak,
the mass density profiles 
vary less dramatically as $r \to 0$ than those in Figure 3
with large central~peaks.

\subsection{NGC 2403, NGC 3198}

With their rotation curves being classified as ``rigid-body type''
\cite{sofue99},
{\em NGC 2403} and {\em NGC 3198} have rotation velocities
increasing gradually from the galactic center almost like 
rigid-body rotation for a considerable radial distance 
before leveling off.
Figure 5 shows the rotation curves 
(from the Sofue website \cite{sofuewebsite})
and the corresponding mass density profiles of 
{\em NGC 2403} (thick line, which also has the nonzero velocity at $r = 0$
replaced with $V(0) = 0$) and {\em NGC 3198} (thin line),
with $r$ measured in units of $R_g = 19.70$ and $31.05$ (kpc) and 
rotation velocity $V(r)$ in units of $V_0 = 130$ and $160$ (km/s)
for {\em NGC 2403} and {\em NGC 3198}, respectively.
The peak density at $r = 0$ in Figure 5 is further reduced from that 
in Figure 4, due to the less steep change in the rotation velocity 
around the galactic center.

It is noteworthy that the {\em NGC 3198} rotation curve has 
a small spike near $r = 0$, 
which results in a sharp turn in the mass density
around the same location.
Another obvious wiggling spike in the rotation curve
is at $r \sim 0.2$, 
causing a corresponding corner formed in the mass density profile in
that neighborhood. 
Apparently,
the effects of some of the fine features in the rotation curve 
are confined locally in a small nearby neighborhood. 

\begin{figure}[H]
\centering
{\includegraphics[scale=0.55]{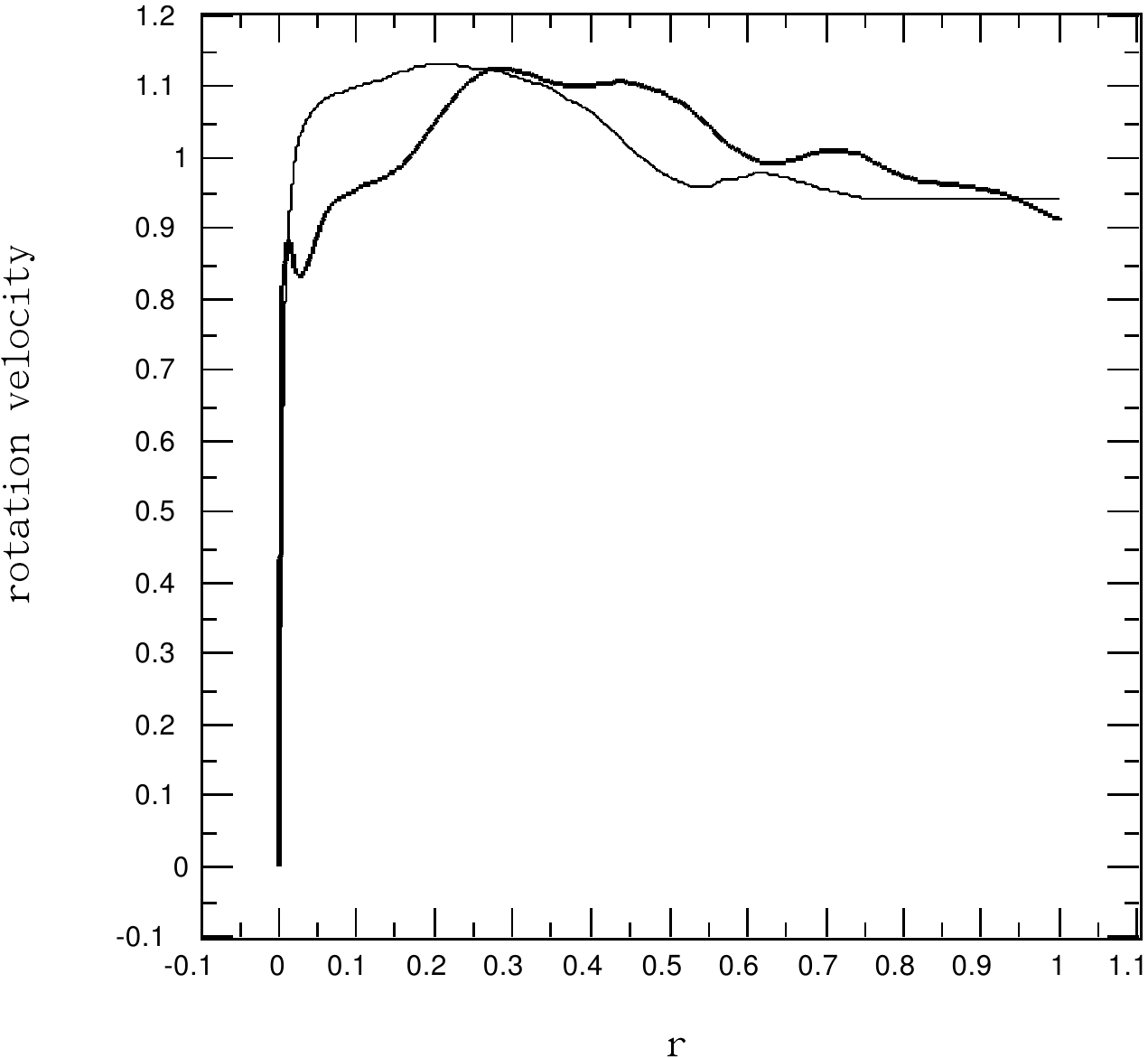}}
{\includegraphics[scale=0.55]{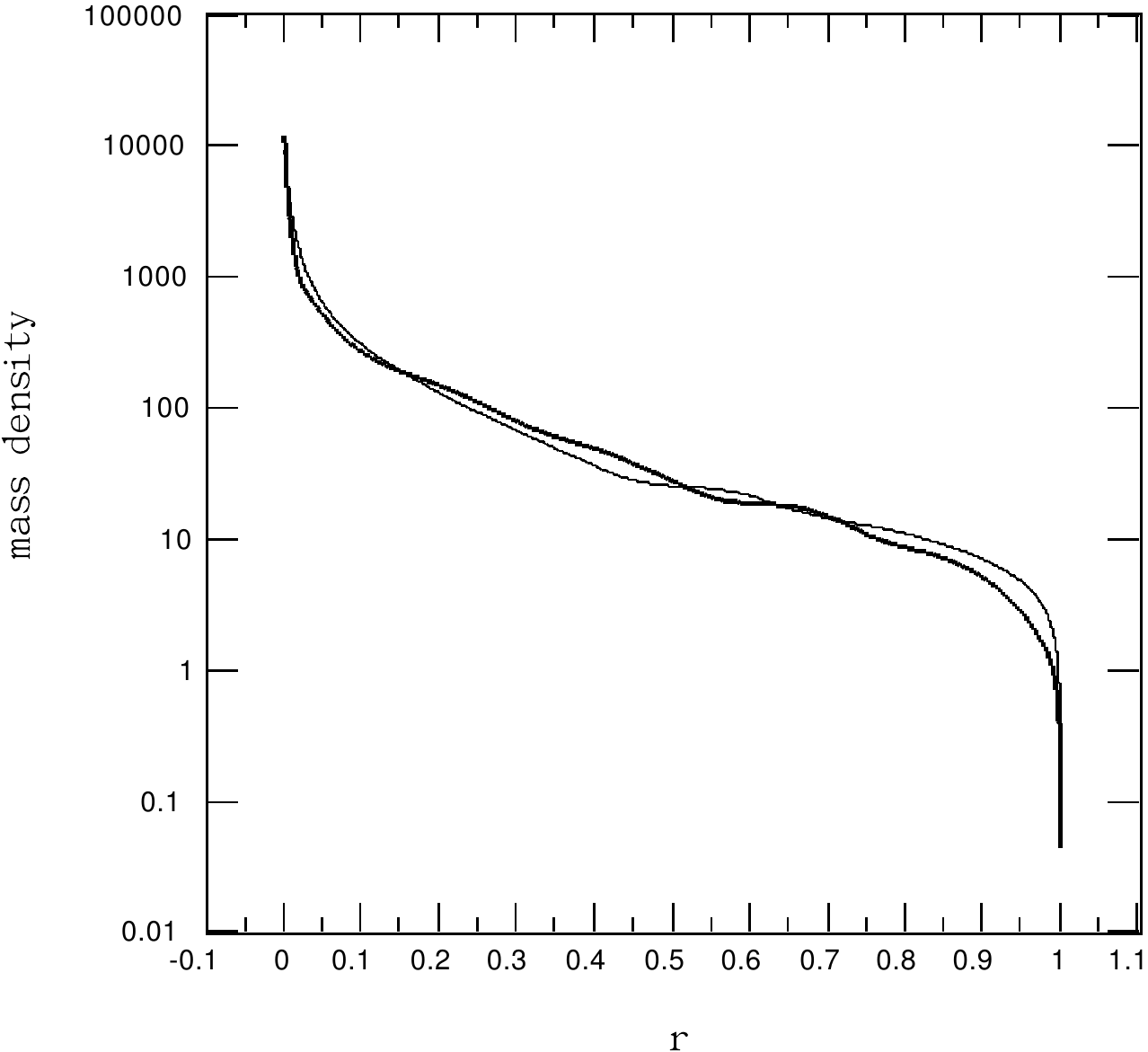}}
\caption{\label{fig:fig4} Profiles of {\em NGC 224} and 
{\em NGC 5055} rotation velocity
$V(r)$ and mass density $\rho(r)$,
with the thick line for that of {\em NGC 224} and the thin line for 
{\em NGC 5055} .
The computed values of the galactic rotation number, $A$,,
are $1.6450$ and $1.6888$ for {\em NGC 224}  and {\em NGC  5055},
respectively.}
\end{figure}

\begin{figure}[H]
\centering
{\includegraphics[scale=0.55]{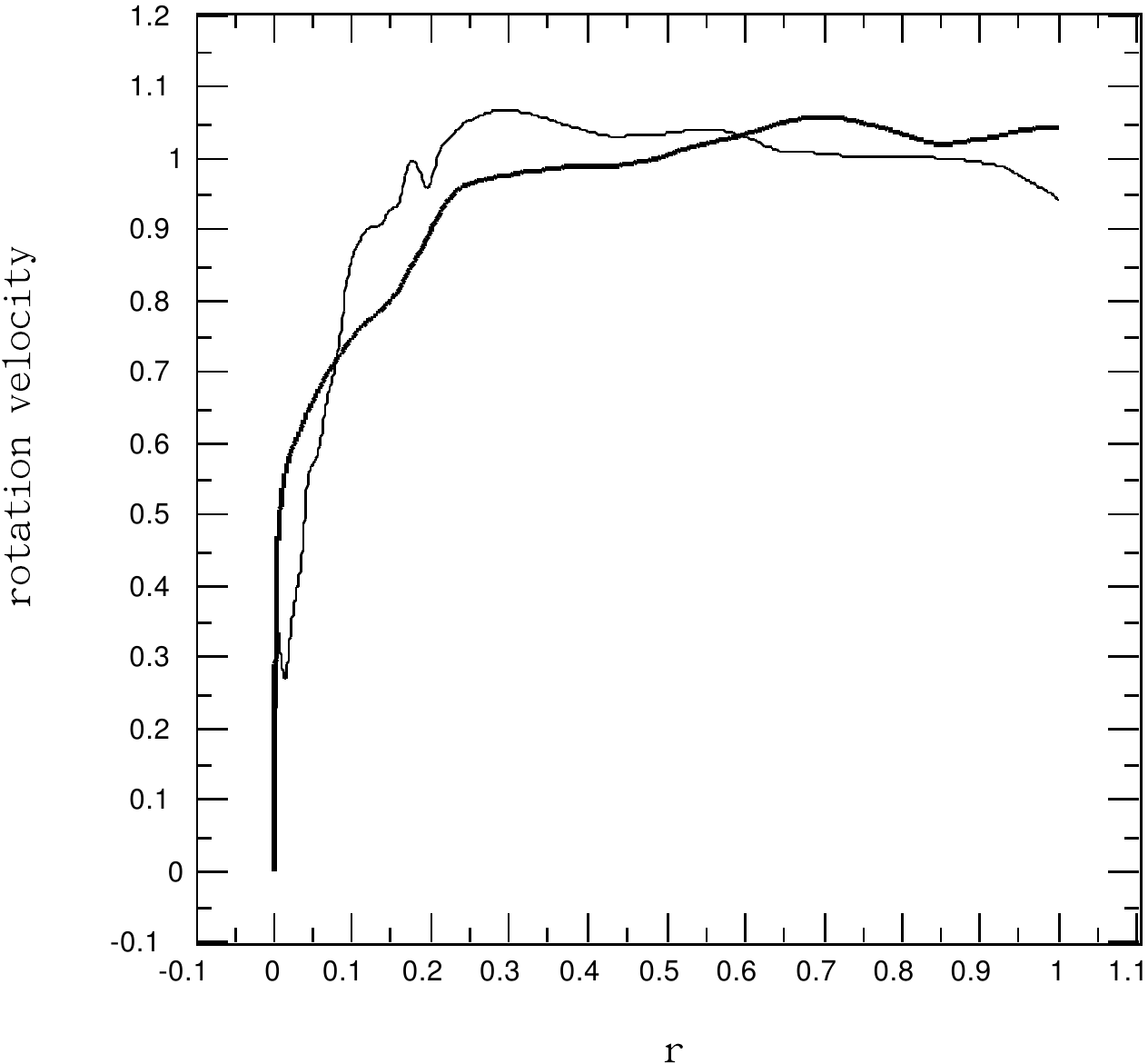}}
{\includegraphics[scale=0.55]{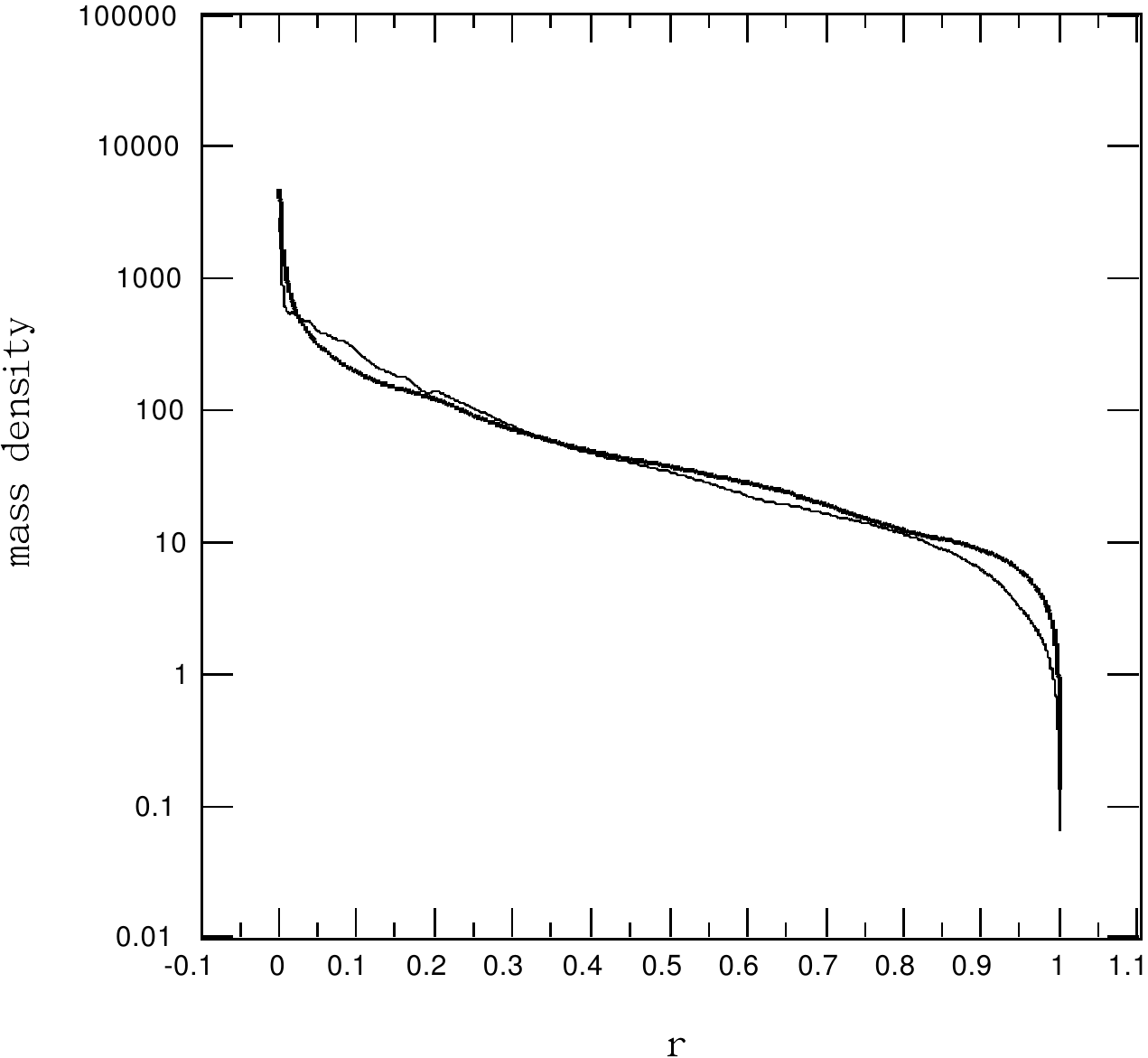}}
\caption{\label{fig:fig5} Profiles of {\em NGC 2403} and 
{\em NGC 3198} rotation velocity
$V(r)$ and mass density $\rho(r)$,
with the thick line for that of {\em NGC 2403} and the thin line for 
{\em NGC 3198}.
The computed values of the galactic rotation number, $A$,
are $1.4918$ and $1.6022$ for {\em NGC 2403} and {\em NGC 3198},
respectively.}
\end{figure}

\newpage
\subsection{Tabulated Summary}

The values of $R_g$ and $V_0$, and the corresponding $A$ and $M_d$, 
for various types of galaxies computed in this section 
are summarized in Table 1 for 
convenient reference. Here, the value of $R_g$ is well-defined as the 
cut-off radius from the rotation curve data, 
such that our non-dimensionalized computational domain 
is always $[0, 1]$, whereas that of the characteristic velocity $V_0$ 
can be somewhat arbitrary and is chosen to approximately reflect 
the average rotation velocity outside the region around galactic center,
which is often fairly flat. 
Thus, the value of computed $A$ can vary depending on the chosen value 
of $V_0$, but the dimensional value of $M_d$ does not change, independent of how $V_0$ is defined.

\begin{table}[H]
\centering
\caption{Values of $R_g$ and $V_0$, and the corresponding $A$ and $M_d$,
for galaxies computed in this~section.}
\centering
\begin{tabular}{cccccc}
\toprule
{\textbf{Galaxy}} & {\textbf{Type}} & {\boldmath$R_g$ \textbf{(kpc)}} & {\boldmath$V_0$ \textbf{(km/s)}} & {\boldmath$A$} & {\boldmath$M_d$ \textbf{(}\boldmath$M_{\odot}$\textbf{)}} \\
\midrule
{\em NGC 4736} (THINGS) & Sab
 & {10.35} & {150} & {1.9656} & {2.756 $\times\,10^{10}$} \\
{\em NGC 4736} (Sofue) & Sab & {10.35} & {150} & {1.5908} & {3.405 $\times \,10^{10}$} \\
Milky Way & Sb
 & {20.55} & {220} & {1.6365} & {1.4138 $\times\, 10^{11}$} \\
{\em NGC 4945} & Sb & {20.00} & {180} & {1.6873} & {8.9337 $\times\, 10^{10}$} \\
{\em NGC 224} & Sb & {31.25} & {250} & {1.6450} & {2.7619 $\times \,10^{11}$} \\
{\em NGC 5055} & Sb & {39.35} & {190} & {1.6888} & {1.9567 $\times\, 10^{11}$} \\
{\em NGC 2403} & Sc & {19.70} & {130} & {1.4918} & {5.1915 $\times \,10^{10}$} \\
{\em NGC 3198} & Sc & {31.05} & {160} & {1.6022} & {1.1541 $\times \,10^{11}$} \\
\bottomrule
\end{tabular}
\end{table}

\section{Discussion}

It should be noted that
the axisymmetric thin-disk continuum model considered here
is at best a simplified approximation of mature spiral galaxies. 
By ``mature'', we mean that the basic galactic structures do not change
(or evolve) with time drastically any more, and thus, their dynamic behavior 
almost approaches the so-called steady state. 
Realistically, however, even those mature spiral galaxies are not 
in a perfect axisymmetric steady state, because they typically exhibit 
spiral arms, tidal streams, bars, \emph{etc}.
\cite{binney87, dehnen98}.
Therefore, a hierarchy of models of increasing sophistication exists 
\cite{binney11}.
However, more sophisticated models usually involve more assumptions,
many of which cannot easily be validated by reliable observational measurements
(such as the dark matter, dark halo, \emph{etc}.) and 
thus become~debatable.
 
On the other hand, we do not think the detailed galactic structures, such as 
spiral arms, \emph{etc}., should alter the averaged gross galactic properties 
in any significant manner.
For example, all the mass density profiles predicted by 
our model in Section 3 exhibit a nearly linear decline (in semi-log plots) if 
the two abruptly varying ends (around $r = 0$ and $r = 1$)
are trimmed out, indicating that
the mass density of most mature galaxies follows approximately a common 
exponential law of decay as typically observed with luminosity measurements. 
However, the radial scale length for the surface mass density of 
many galaxies may not closely
match that of the measured brightness distribution \cite{feng11}.
Yet, for some galaxies, such as {\em NGC 4736}, our predicted mass distribution
(which is basically the same as that by 
Jalocha, Bratek and Kutschera \cite{jalocha08}) has been shown to 
be quite consistent with the $I$-band luminosity profile, 
having a mean mass-to-light ratio $M/L_I$ $= 1.2$ \cite{jalocha08}.
Because there are no well-established theoretical constraints for 
the mass-to-light ratio, 
quantitative discussion on the discrepancy between our predicted 
mass distribution and the measured brightness distribution
can hardly avoid lack of 
the desired scientific rigor and, therefore, is not pursued further 
in the present work. 
Here, we choose to focus on the axisymmetric thin-disk model 
with minimum physical assumptions, and apply only the well-established 
Newtonian dynamics to determine the mass distribution in rotating spiral galaxies 
based on observed rotation curves.
Along with appropriate mathematical treatments, we believe our results 
provide logically rigorous references for 
more sophisticated model development.
Hence, in this section, we first discuss a few extensions from
the strict thin-disk model results presented in the previous section.
Other relevant topics, such as the applicability of Keplerian dynamics and 
circular orbit stability, are also examined.

\subsection{Nonzero Rotation Velocity at $r = 0$}

In Section 3.2, 
we have treated the rotation curves having nonzero velocity
at $r = 0$ by replacing the value of $V(0)$ with a zero value 
in the rotation curve data file.
Such a simplistic approach may be a little distasteful 
to some people with a rigorous mind.
Therefore, a more elaborated treatment is \mbox{provided here.}

With the thin-disk model, we have demonstrated with 
galaxies of various types of realistic rotation curves 
that the mass density is always highest at the galactic center,
and a nonzero rotation velocity at $r = 0$ corresponds to 
an infinite mass density at the galactic center.
To enable the numerical treatment of the infinite local mass density,
it may not be unreasonable to consider the galaxies with 
nonzero rotation velocity at $r = 0$ to consist of 
a dense spherical core at the galactic center in addition to 
a self-gravitating thin disk.
In that case, we should modify (\ref{eq:force-balance}) to 
include a term due to the dense core with a spherically symmetric 
gravitational field. 
Among many choices, we can simply assume a spherical core confined within 
a small volume, e.g., in $r < R_c$ $= 0.0001$, having a mass 
$M(r) = A \, V(0)^2 \, r$, where $V(0)$ is nonzero according to the measured
rotation curve. 
This corresponds to a spherically symmetric mass density 
$\rho(r) = [d M(r)/d r] / (2 \pi r^2)$ $= A \, V(0)^2 / (2 \pi r^2)$ 
in $r < R_c$, becoming infinite as $r \to 0$.
As a consequence, the second term in 
Equation~(\ref{eq:force-balance}), namely $\frac12 A \, V(r)^2$, 
can be replaced by
$\frac12 A \, [V(r)^2 - V(0)^2]$ for $r < R_c$ and by 
$\frac12 A \, [V(r)^2 - V(0)^2 R_c/r]$ for $r \ge R_c$. 
Such a modification is actually equivalent to 
replacing the original rotation curve, $V(r)$, with a modified 
one that becomes zero at $r = 0$ as:
\begin{eqnarray} \label{eq:modified_V}
\left\{
\begin{array}{cc}
\sqrt{V(r)^2 - V(0)^2} \, , \quad r < R_c 
\\
\sqrt{V(r)^2 - V(0)^2 R_c/r} \, , \quad r \ge R_c 
\end{array} \right . \, 
\end{eqnarray}

If we apply this approach to the Milky Way, which has 
$V(0) = 0.9282$, 
with $R_c = 10^{-4}$, we obtain $A = 1.6368$ (instead of $1.6365$ in Section 3.2).
Thus, the total mass in the Galactic disk is 
\mbox{$M_d = 1.4135 \times 10^{11}$ $M_{\odot}$ }
(instead of $1.4138 \times 10^{11}$ $M_{\odot}$ in Section 3.2).
The mass in the spherical core is $A \, V(0)^2 \, R_c \, M_d$ 
$= 1.6368 \times 0.9282^2 \times 1.4135 \times 10^7$
$= 1.9933 \times 10^7$ $M_{\odot}$.
The combined mass of the core and disk is then 
$1.4137 \times 10^{11}$ $M_{\odot}$, 
which is of no practical difference from the value in Section 3.2. 
With such a small core of $R_c = 10^{-4}$, the modified rotation curve 
Equation~(\ref{eq:modified_V}) is also of no practical difference from
that (thick line) in Figure 3, having $<$2\% change at $r = 0.0024$ 
(the second data point in measured rotation curve),
$<$1\% change at $r = 0.0049$ (the third data point),
$\sim$0.5\% change at $r = 0.0073$ (the forth data point), 
and so on and so forth.

However, if we assume $R_c = 0.01$ for a bigger core, 
the computed value of $A$ 
for the Milky Way becomes
$1.6599$ and the corresponding mass in the galactic disk then is
$M_d = 1.3939 \times 10^{11}$ $M_{\odot}$.
Combining with the mass of the core
($A\, V(0)^2 \, R_c \, M_d$ $= 1.9934 \times 10^9$ $M_{\odot}$),
we have a total galactic mass of
$1.4138 \times 10^{11}$ $M_{\odot}$, 
which is basically the same as that in Section 3.2.
Hence, the total galactic mass remains unchanged 
for a substantial range of the spherical core size, $R_c$.
However with $R_c = 0.01$, the modified Milky Way disk rotation curve
according to Equation~(\ref{eq:modified_V}) differs noticeably from
the original one provided by Sofue (as shown in Figure 6),
especially around the galactic core, where the influence of
the gravitational field of the spherical core is more significant.
Yet, the computed $\rho(r)$ in the thin disk still appears 
indistinguishable from that in Figure 3, except that the peak value 
at $r = 0$ is reduced to $3650$ from $25,262$ (in Figure 3).
This is because in a small core at the galactic center 
the details of mass distribution, whether axisymmetrically or spherical
symmetrically, cannot make much of a difference in the gravitational field
some distance away.

\begin{figure}[H]
\centering
{\includegraphics[scale=0.60]{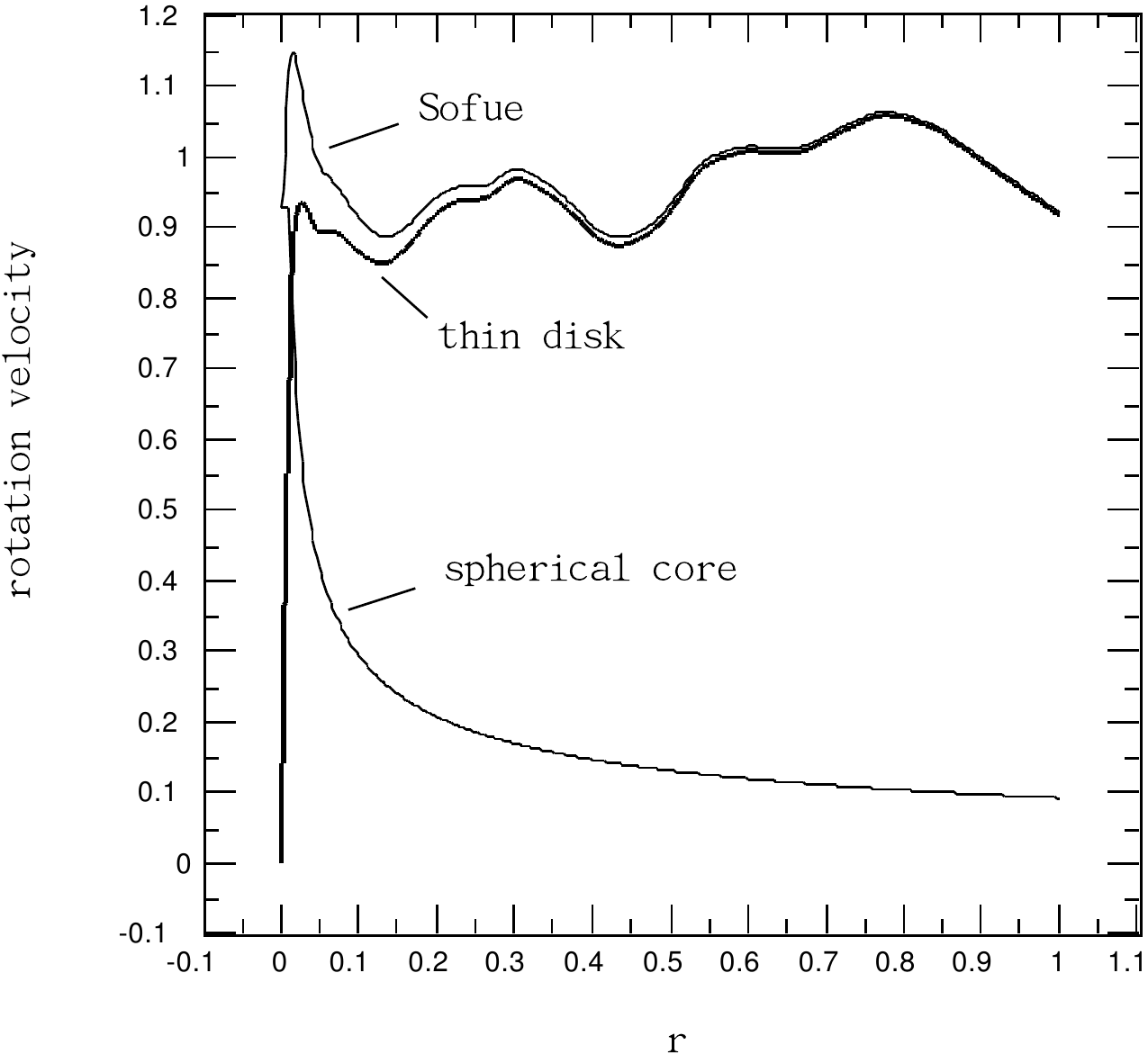}}
\caption{\label{fig:fig6} Profiles of the Milky Way rotation curves 
when decomposed into that corresponding to 
a spherical core of $R_c = 0.01$ and
a thin disk. The original rotation curve from Sofue is also shown
here as a reference.}
\end{figure}

It seems though that the effort of decomposing the galaxy 
into a small spherical core and 
a thin disk only helps treat rotation curves with 
nonzero velocity 
at $r = 0$ in a mathematically elegant manner,
such that the need of explicitly considering 
the infinite mass density is eliminated.
Physically, a nonzero rotation velocity at $r = 0$ has 
unclear meanings and should remain as a debatable subject;
so should the implication of the corresponding infinite mass density,
because the common wisdom usually indicates that 
``nature abhors infinities''.
Thus, we prefer the straightforward treatment in Section 3.2 
to simply bring the rotation velocity to zero at $r = 0$, 
especially when it does not seems to be, at 
the expense of compromising the general result's accuracy.
The insensitivity of mass distribution in the galactic disk 
and total galactic mass
to detailed descriptions of the structure in 
a small spherical central core
illustrated here is
consistent with the findings of previous authors
({\em cf.} the discussion of Nordsieck \cite{nordsieck73} and the citations therein).

\subsection{Central Bulge in Disk Galaxy}

Yet, our methodology for treating a central spherical core
can be easily extended for analyzing galaxies with 
a considerably larger central bulge with 
{\em a priori} given spherical mass distributions.
As an example, assuming the Milky Way rotation curve
in Figure 3 to be a result of the combination of a central bulge with
a spherically symmetric mass density:
\begin{equation} \label{eq:bulge-density}
\rho_b(r) = \rho_{b0} e^{-(r/R_b)^3} 
\, \, 
\end{equation}
and an axisymmetrically distributed mass in a thin-disk,
Equation~(\ref{eq:force-balance}), with $V(r)$ in Figure 3 being replaced by:
\begin{equation} \label{eq:bulge-modified-V}
\sqrt{V(r)^2 - \frac{4\pi\,\rho_{b0}\,R_b^3}{3\,A} 
\left[1 - e^{-(r/R_b)^3}\right]} 
\, \, 
\end{equation}

For $R_b = 0.2$ and $\rho_{b0}/A = 7$, the disk rotation curve 
as determined from Equation~(\ref{eq:bulge-modified-V}) is shown in
Figure 7 together with the computed disk mass density distribution. 
In the presence of this bulge, the value of $A$ becomes $2.4793$, 
and the disk mass density exhibits a dip around $r = 0.12$.
Thus, \linebreak$\rho_{b0} = 7 \times 2.4793 = 17.3551$, 
and the corresponding bulge density profile
is also shown in Figure~7.
The mass in the disk portion (calculated from Equation~(\ref{eq:M_d})) is 
$M_d = 9.3320 \times 10^{10}$ $M_{\odot}$
and that in the bulge portion
$M_b = 4\pi\,\rho_{b0}\,R_b^3 M_d/3$ $= 5.4273 \times 10^{10}$ $M_{\odot}$.
The total galactic mass $M_g = M_d + M_b$ 
$= 1.4759 \times 10^{11}$ $M_{\odot}$
(instead of $1.4138 \times 10^{11}$ $M_{\odot}$ predicted by a 
pure disk model in Section 3.2).
Because a substantial amount of the mass is concentrated in the central bulge
with its portion of spherically symmetric mass density practically diminishing 
for $r > 0.3$, 
the disk surface mass density in the solar neighborhood 
around $r = 0.3893$ (corresponding to $8$ kpc) 
becomes $\rho(0.3893)\,M_d\,h/R_g^2$ $=108$ $M_{\odot}$/pc$^2$)
where $\rho(0.3893) \sim 49$ from the ``disk'' mass density curve 
in Figure 7.
Even though the presence of our example bulge causes 
only a few percent of increase in the 
total galactic mass from that predicted by a pure disk model,
the disk surface mass density in the solar neighborhood 
can decrease by 25\%. 

\begin{figure}[H]
\centering
{\includegraphics[scale=0.55]{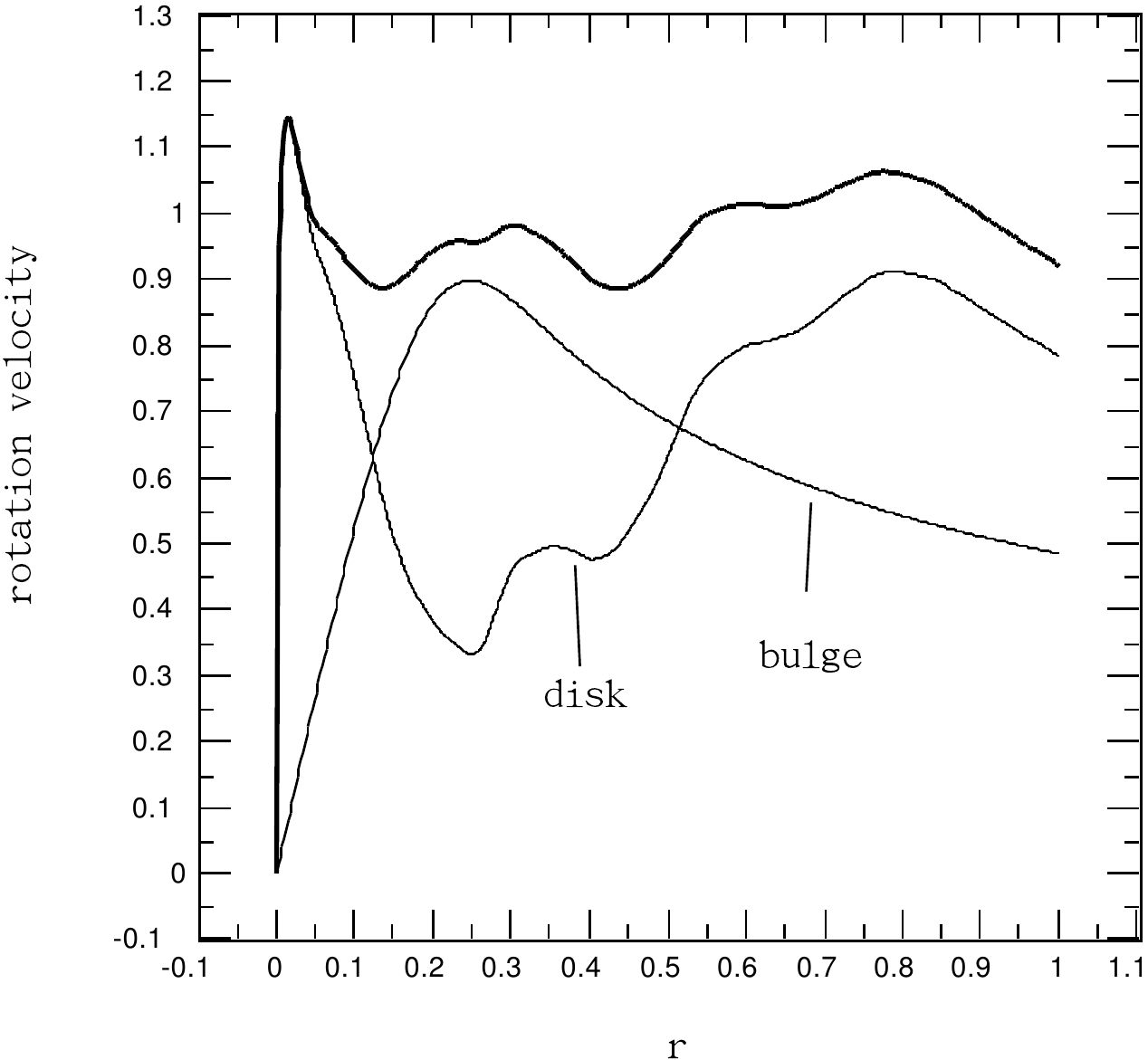}}
{\includegraphics[scale=0.55]{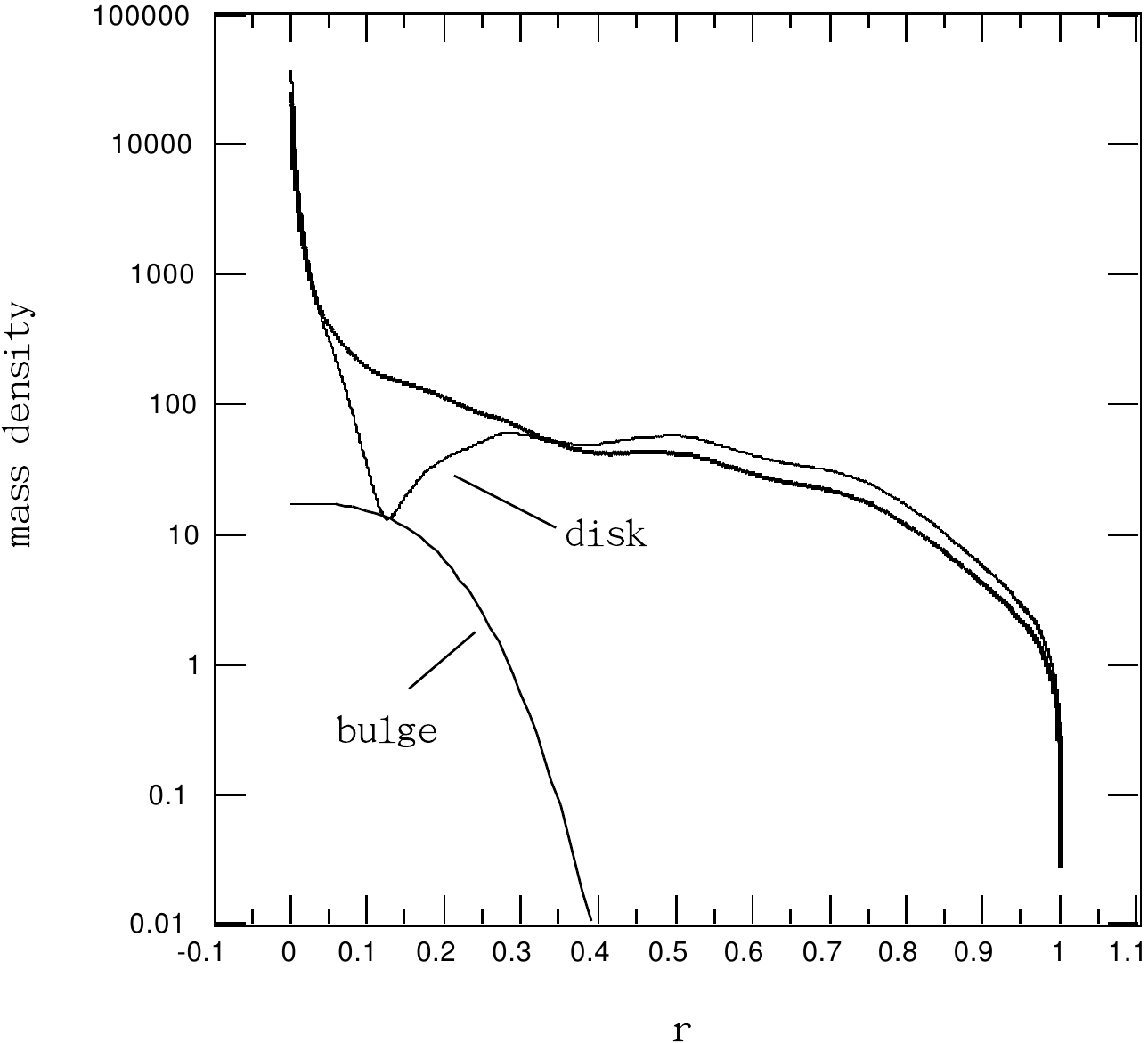}}
\caption{\label{fig:fig7} Profiles of the Milky Way 
rotation velocity
and mass density for the disk portion and bulge portion 
(with $R_b = 0.2$ and $\rho_{b0}/A = 7$) as noted 
along with that in Figure 3 (thick line) as a reference.}
\end{figure}

If $R_b = 0.25$ and $\rho_{b0}/A = 5$,
the value of $A$ will become $3.0223$, and therefore, $\rho_{b0} = 15.1115$.
As a consequence, 
$M_d = 7.6554 \times 10^{10}$ $M_{\odot}$, 
$M_b = 7.5716 \times 10^{10}$ $M_{\odot}$
and 
$M_g = 1.5227 \times 10^{11}$ $M_{\odot}$.
The value of
$\rho(0.3893)$ is $\sim$41 corresponding to 
the disk surface mass density in the solar neighborhood of
$74$ $M_{\odot}$/pc$^2$.
Thus, for a given rotation curve, 
the actual value of disk surface mass density in a galaxy
can vary significantly when considering a model with 
a combination of a spherical bulge and an axisymmetric thin disk,
depending upon the bulge mass structure. 
The bulge mass structure described by Equation~(\ref{eq:bulge-density}) 
is only for illustrative purpose 
with the convenience of mathematical manipulation. 
The fact that adding a spherical bulge offers 
a another degree of freedom for adjusting mass distribution in 
the galactic disk should not depend upon 
the details in the bulge mass structure.
This extra degree of freedom comes at the expense of uncertainty 
due to the difficulties in
determining the bulge mass structure that 
is governed by much more complicated physical processes than
simply balancing the gravitational force and 
centrifugal force.
Hence, we choose to take the bulge mass structure as given {\em a priori} 
in analyzing mass distribution in disk galaxies according to
Newtonian dynamics, 
with our focus kept on the thin-disk portion of galaxies.

Without considering the central bulge, the mass distribution in
the galactic disk can be uniquely determined from a given 
rotation curve. 
With the central bulge, its mass structure must be known {\em a priori}
in order to compute a unique disk mass distribution 
corresponding to the given rotation curve.
However, how to reliably determine the bulge mass structure 
besides using its luminosity information 
and an assumed mass-to-light ratio seems to be 
an open question.

Nevertheless, our illustrative analysis presented here 
demonstrates the general effect of a central bulge to
basically shift mass from the periphery toward the center of 
a galaxy for a given rotation curve. 
The more massive a central bulge becomes, 
the less mass is needed in the disk periphery region
according to Newtonian dynamics.
Yet, the total mass in a galaxy seems to be much less sensitive 
to the presence or absence of a central bulge.

\subsection{Rotation Velocity beyond the Cut-off Radius}

Furthermore, as shown in Section 3.1, 
our finite-disk galaxy model and the associated computational method 
can further be used to determine 
the rotation velocity of 
matters outside the cut-off radius,
which we assume to be the edge of galaxy 
where the mass density diminishes.
Again, taking the Milky Way as an example,
Figure 8 shows the computed rotation velocity 
beyond the galactic edge $r = 1$, 
as a continuation from the measured rotation curve 
that ends at $r = 1$ 
and gradually approaching the Keplerian rotation curve 
for $r > 2$.
Here, the Keplerian rotation curve is generated by applying
\mbox{Keplerian dynamics:}

\begin{equation} \label{eq:keplerian}
V_K(r) =
\sqrt{\frac{2\pi}{r \, A} \int_0^r \rho(\hat{r}) \, h \, \hat{r} d\hat{r}}
\, \, 
\end{equation}
with $\rho(\hat{r})$ and $A$ being obtained through computations in Section 3.2.
Because Keplerian dynamics cannot correctly describe 
the situation of disk galaxies with a non-spherically symmetric 
gravitational field,
the rotation curve predicted by Keplerian dynamics Equation~(\ref{eq:keplerian})
from the disk mass distribution, $\rho(\hat{r})$, 
differs noticeably from that of Newtonian dynamics (as depicted 
with the thick line and its extension in Figure~8).
Only at a large distance (e.g., $r > 2$) from 
the galactic disk does the Keplerian rotation curve 
approach that computed based on Newtonian dynamics,
for the effect of the disk structure diminishes at a large distance, 
where the gravitation field of a finite disk galaxy approaches 
the spherically symmetric field of a point~mass. 
\begin{figure}[H]
\centering
{\includegraphics[scale=0.60]{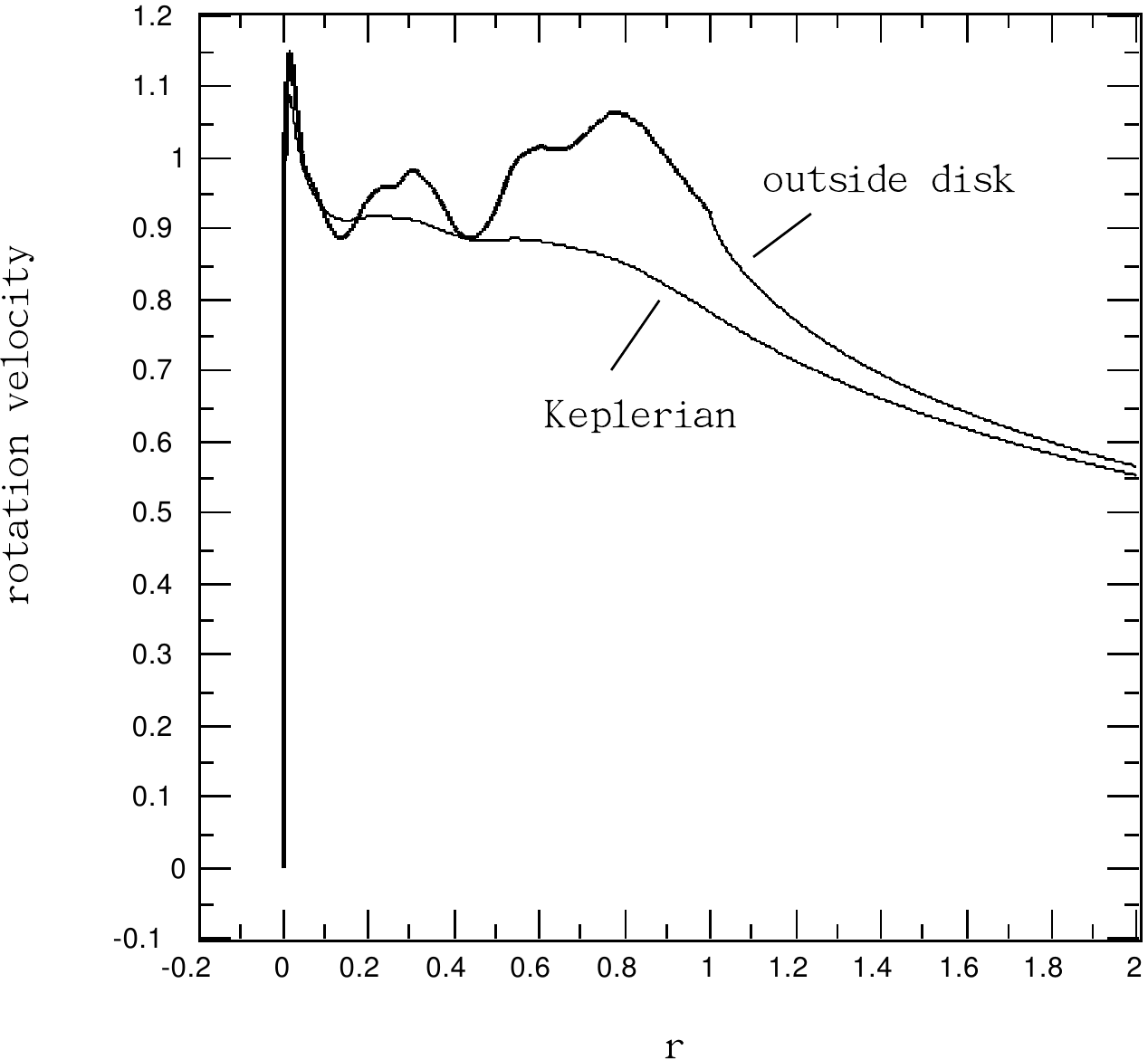}}
\caption{\label{fig:fig8} Profiles of the Milky Way rotation curves 
of the originally measured one (with imposed $V(0) = 0$, thick line)
and its extension outside the disk edge at $r = 1$ (the thin line as labeled),
as well as $V_K(r)$ according to Keplerian formula 
Equation~(\ref{eq:keplerian}) based on 
the mass density, $\rho(r)$, shown in Figure 3 (the thin line as labeled).}
\end{figure}

\subsection{Applicability of Keplerian Dynamics}

If Keplerian dynamics were applied to estimate the amount of mass 
within the solar radius (8 kpc corresponding to $r = 0.3893$)
from the measured local rotation velocity (201.0658 km/s corresponding to
$V(0.3893) = 0.9139$),
we would obtain $M_K(r) = A\, r\, V(r)^2$ 
$= 1.6365 \times 0.3893 \times 0.9139^2$ $= 0.5321$, 
which corresponds to $0.5321 \times 1.4138 \times 10^{11}$ 
$= 0.7523 \times 10^{11}$ $M_{\odot}$.
The actual amount of mass within the solar radius calculated using 
$M(r) = 2\pi\,h \int_0^r \rho(\hat{r})\, \hat{r} d\hat{r}$ at $r = 0.3893$
based on the computed $\rho(r)$ in Section 3.2, 
is $0.5078$, which corresponds to $0.7179 \times 10^{11}$ $M_{\odot}$.
Although the mass within the solar radius ($r = 0.3893$)
estimated with Keplerian dynamics ($M_K(0.3893)$ $= 0.5321$)
does not seem too far off the actual value ($M(0.3893)$ $= 0.5078$),
the value of $M_K(r)$ deviates more and more from $M(r)$ with 
increasing $r$, as can be seen in Figure 9.
The value of $M_K(r)$ may even decrease with $r$,
when calculated 
according to the measured rotation curve (as clearly shown in Figure 8
for $r > 0.85$).
Because $M_K(r)$ is expected to monotonically increase with $r$, 
for there is no physical evidence of negative mass in the universe,
a negative slope of $M_K(r)$ \emph{versus} $r$ indicates a failure of
Keplerian dynamics for correctly predicting 
the mass distribution corresponding to 
the measured rotation curve for the Milky Way, even in a qualitative sense; 
or, in other words, 
a rotation curve that does not satisfy $d[r\,V(r)^2]/dr \ge 0$, namely:

\begin{equation} \label{eq:sphericity}
\frac{dV(r)}{dr} \ge - \frac{V(r)}{2r}
\, \, 
\end{equation}
is inconsistent with spherically symmetric gravitational potential,
and thus, Keplerian dynamics becomes inapplicable in a strict sense.

\begin{figure}[H]
\centering
{\includegraphics[scale=0.60]{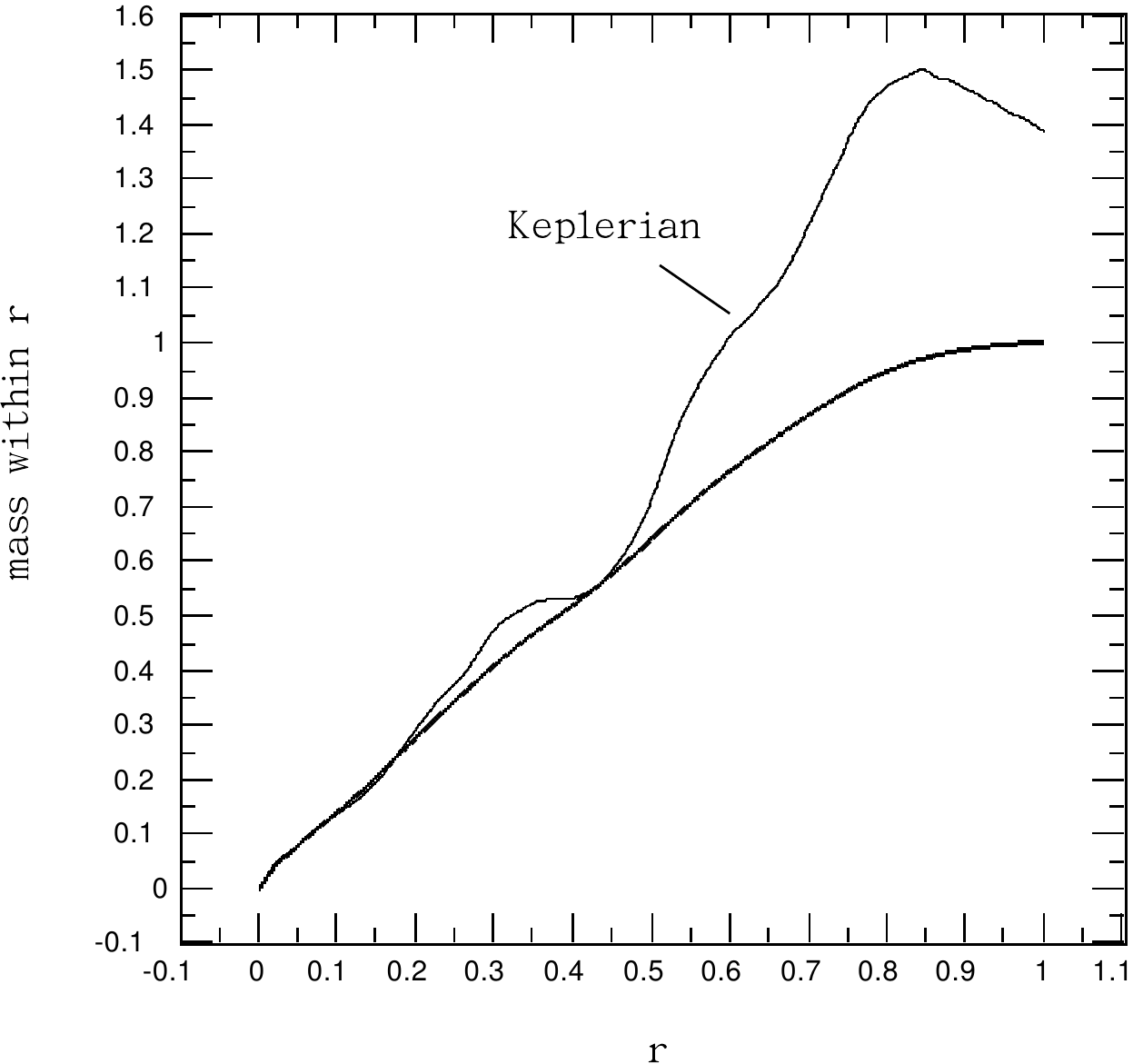}}
\caption{\label{fig:fig9} Profiles of the Milky Way mass within radial 
distance $r$ (thick line), 
$M(r) = 2\pi\,h \int_0^r \rho(\hat{r})\, \hat{r} d\hat{r}$, 
and that estimated with Keplerian dynamics (thin line as labeled),
$M_K(r) = A\, r\, V(r)^2$.} 
\end{figure}
The same condition of Equation~(\ref{eq:sphericity}) was referred to as
the sphericity condition by 
Jalocha, Bratek and Kutschera \cite{jalocha08}, 
Bratek, Jalocha and Kutschera \cite{bratek08} and 
Jalocha \emph{et al}. \cite{jalocha10} 
and the violation of which was used as an indication of 
the disk model being more appropriate for determining the mass distribution
and the presence of a massive spherical halo of non-baryonic dark matter
being unlikely. 
Actually, (\ref{eq:sphericity}) is only a necessary condition 
for the sphericity of gravitational potential to exist,
but not sufficient. 
A rotation curve satisfying (\ref{eq:sphericity}) does not 
guarantee that it must correspond to a spherically symmetric gravitational
field.
Feng and Gallo
\cite{feng11} showed that a flat rotation curve can be described by 
both a spherically symmetric and an axisymmetric disk mass distribution.
However, using a spherically symmetric mass model, namely Keplerian dynamics, 
to describe a rotating disk galaxy can lead to erroneous results
and conclusions.

\subsection{Circular Orbit Stability}

It is interesting to note that the mathematical form of 
the sphericity condition Equation~(\ref{eq:sphericity}) 
appears quite similar to the necessary condition for 
circular orbit stability: 

\begin{equation} \label{eq:stability}
\frac{dV(r)}{dr} \ge - \frac{V(r)}{r}
\, \, 
\end{equation}
which can be derived from the consideration of angular momentum 
conservation for a rotating object slightly deviating from its 
original (circular) orbit as follows. 
An object that is rotating with a velocity, $V(r)$, at radial coordinate $r$
possesses an angular momentum, $r\,V(r)$.
If it deviates from its original orbit at $r$ to $r + \delta r$
(due to some sort of perturbations), 
its rotation velocity should change from $V$ to $V + \delta V$, such that
$(r + \delta r)$ $(V + \delta V)$ $= r\,V$ or 
$\delta V/\delta r = -V/r$ ($\le 0$, \emph{i.e.}, $\delta V < 0$ 
when $\delta r > 0$ and $\delta V > 0$ when $\delta r < 0$), according to 
the conservation of angular momentum.
On the other hand, this object is subjected to a 
gravitational force at $r + \delta r$ 
equal to $m V(r + \delta r)^2/(r + \delta r)$,
where $m$ is its mass and $V(r + \delta r)$ is the rotating velocity
of objects at $r + \delta r$ according to the rotation curve. 
Thus, for this object to be pulled back by gravitational force 
to its original orbit,
namely for its orbit to be centrifugally stable,
we must have $V + \delta V < V(r + \delta r)$ for $\delta r > 0$
and $V + \delta V > V(r + \delta r)$ for $\delta r < 0$.
This leads to 
\mbox{$\delta V$ $< \delta r\, dV(r)/dr + o(\delta r^2)$} for $\delta r > 0$
and $\delta V$ $> \delta r\, dV(r)/dr + o(\delta r^2)$ for $\delta r < 0$
as a result of Taylor expansion around $\delta r = 0$,
namely $\delta V / \delta r = -V/r < dV/dr$ as $\delta r \to 0$.

In the case of the solar system with a point mass at $r = 0$,
the planets rotate following the Keplerian rotation curve 
$dV/dr = -V/(2r)$ (taking the equal sign in Equation~(\ref{eq:sphericity})
for the mass, $M_K(r)$, does not change for $r > 0$).
Because $-1/2 > -1$, the Keplerian rotation curve satisfies
the circular orbit stability condition Equation~(\ref{eq:stability}), 
as evidenced by the existence of the solar system with many planets 
circling around the Sun year after year. 

Many spiral galaxies exhibit nearly 
flat rotation curves ({\em cf.} the review of Sofue and Rubin 
\cite{sofue01}),
corresponding to $dV/dr \sim 0$, which can easily 
satisfy Equation~(\ref{eq:stability}).
Thus, the rotating matter distributed in circular orbits of the galactic disk,
as can be computed with the method illustrated in the present work, 
for flat rotation curves are stable 
in the sense similar to that of the planets circling around the Sun.
However, Equation~(\ref{eq:stability}) is only a necessary condition for stability.
There have been many other (necessary) conditions proposed in the literature
for rotating disk galaxy stability, which often seem controversial, 
as critically discussed by Jalocha \emph{et al}. \cite{jalocha10}. 
The circular orbit stability condition Equation~(\ref{eq:stability}) 
derived in the present work is established from a concrete physical principle
and can be used to examine the validity of measured rotation curves.
Especially for those rotation curves containing decreasing velocity
at a large radius, the stability condition Equation~(\ref{eq:stability})
with its right side $\propto-1/r$
is likely violated.
The portion of a rotation curve not satisfying Equation~(\ref{eq:stability})
may point to the issues with too serious deviations from circular orbits 
and axisymmetry due to the spiral arms.
After all, the axisymmetric disk model with rotation velocity depending
only on radius
is a tremendous simplification 
of a realistic rotating galaxy; 
such a simplified description of reality should be 
regularly checked \mbox{for consistency.} 

For the Milky Way rotation curve ({\em cf.} Figure 3),
the portion of $r > 0.8$ has $dV/dr \sim -0.75$, 
while $0.9 < V/r < 1.32$.
Thus, we have $-0.75 > -0.9$, which satisfies 
the circular orbit stability condition Equation~(\ref{eq:stability}),
but not the sphericity condition (\ref{eq:sphericity}),
which is consistent with that shown in Figure 9.
Thus, the Milky Way appears to be appropriately described 
with the thin-disk model or with a combination of 
a central bulge and a thin disk.

For the {\em NGC 4736} rotation curve of Sofue ({\em cf.} Figure 1),
the negative slope in the outer region $r > 0.6$ is 
$dV/dr \sim -0.875$, while $0.85 < V/r < 2$.
Thus, circular orbit instability is likely to occur 
in the outer region of {\em NGC 4736} if the rotating matter indeed follows
the rotation curve of Sofue. 
In a relative sense, 
the THINGS version of the {\em NGC 4736} rotation curve \citep{deblok08}
has a less steep negative slope than that of Sofue,
indicating that the THINGS version describes a more stable 
circular motion of rotating matter.

\section{Conclusions}

With the computational method presented here, 
mass distributions in mature spiral galaxies corresponding to 
various types of measured rotation curves can be 
accurately determined
by solving a linear algebra matrix equation,
which clarifies the uniqueness of the solution 
when it exists.
Our formulation for the finite thin-disk model is
based solely on Newtonian dynamics without the need of fictitious 
rotation velocity outside the cut-off radius, 
with a belief that the cut-off radius in the rotation curve measurement
is a consequence of the absence of matter beyond such 
a galactocentric distance.
All the mass density profiles 
predicted by our model for various galaxies exhibit 
approximately a common exponential law of decay
(if the two abruptly varying ends are trimmed out),
qualitatively consistent with typically observed luminosity measurements. Thus, we think Newtonian dynamics can be quite adequate 
for self-consistently describing the rotation behavior 
of mature spiral galaxies.

Despite difficulties in clarifying the physical meaning,
nonzero rotation velocities at the galactic center ($r = 0$) 
were reported in rotation curve measurements for 
several galaxies \cite{sofue99}.
The nonzero value of rotation velocity at $r = 0$
mathematically corresponds to unbounded local mass density in the pure disk model,
which is intractable in numerical computations.
Such a numerical challenge can be avoided by placing a small 
spherical core at $r = 0$.
Thus, the rotation velocity in the galactic disk is modified accordingly
by subtracting out the spherical core effect, and 
the disk mass distribution can be consistently computed from 
the modified rotation curve.
As long as the size of the spherical core is sufficiently small, 
our computed results show that no noticeable
change in the disk mass distribution that can be observed 
while the nonzero velocity at $r = 0$ being elegantly dealt with.

To examine the basic effect of an observed central bulge,
we assume a spherically symmetric mass structure for the bulge, 
whose gravitational effect can be conveniently incorporated in our 
thin-disk formulation.
Our results indicate that the presence of a central bulge 
tends to shift mass from the periphery toward the galactic center 
with little change in the total galactic mass.

Extending the computational domain beyond the galactic edge 
enables us to also compute rotation velocity outside the cut-off radius. 
Beyond the galactic edge where the mass density should be negligible,
the computed rotation velocity does not follow 
the Keplerian profile until out over $r > 2$. 

By applying the principle of angular momentum conservation, 
a necessary condition for circular orbit stability can be derived.
It appears that the galaxies with flat or rising rotation velocities
are more stable than those with declining rotation velocities.
Especially in the region near the galactic edge, those rotation curves 
having too steep of a negative slope are likely to violate 
the condition for circular orbit stability, and 
therefore, their validity for realistically describing 
galactic rotational characteristics may become questionable.

\acknowledgements{Acknowledgments}

We are indebted to Len Gray of Oak Ridge Laboratory
for teaching us detailed boundary element techniques
for the elegant removal of various singularities in 
integral equations.
A conversation with Professor Ming Cai of Florida State University
helped us establish the condition Equation~(\ref{eq:stability}) for 
circular orbit stability.
Several communications with Lukasz Bratek of 
the Polish Academy of Sciences on mathematical details in our 
computational method stimulated our interest 
in examining {\em NGC 4736}, among other galaxies. 
We also want to thank Louis Marmet, Ken Nicholson and 
Professor Michel Mizony for sharing their 
physical intuitions and the results of 
disk galaxy computations.
Some of the reviewers' comments helped enhance 
our presentation and therefore are appreciated. 

\newpage


\authorcontributions{Author Contributions}


In preparing the present article, James Q. Feng took the 
responsibility of deriving mathematical formulations, 
carrying out computational analysis, and writing manuscript,
while C. F. Gallo provided background knowledge, physical insights, 
and interpretations
of results.


\conflictofinterests{Conflicts of Interest}

{The authors declare no conflict of interest. }
     

\bibliographystyle{mdpi}
\makeatletter
\renewcommand\@biblabel[1]{#1. }
\makeatother


\end{document}